\documentclass[12pt]{article}
\usepackage{latexsym,amssymb,amsmath}
\usepackage{amsthm, amstext}
\usepackage{array, amsfonts, mathrsfs}
\usepackage[dvips]{graphicx}

\usepackage{epsf}

\newtheorem{Convention}{Convention}

\numberwithin{equation}{section}

\begin{document}
\date{}
\author{M.I.Belishev\thanks {Saint-Petersburg Department of
                 the Steklov Mathematical Institute, Saint-Petersburg State University, Russia;
                 belishev@pdmi.ras.ru.}\,\,
                 and N.Wada\thanks{Graduate school
                 of Informatics Kyoto University, Japan;
                 naoki@acs.i.kyoto-u.ac.jp}}

\title{A C*-algebra associated with dynamics on a graph of strings}
\maketitle

\begin{abstract}
An operator C*-algebra $\mathfrak E$ associated with a dynamical
system on a metric graph is introduced. The system is governed by
the wave equation and controlled from boundary vertices. Algebra
$\mathfrak E$ is generated by the so-called {\it eikonals}, which
are self-adjoint operators related with reachable sets of the
system. Its structure is the main subject of the paper. We show
that $\mathfrak E$ is a direct sum of "elementary blocks". Each
block is an algebra of operators, which multiply ${\mathbb
R}^n$-valued functions by continuous matrix-valued functions of
special kind. The eikonal algebra is determined by the boundary
inverse data. This shows promise of its possible applications to
inverse problems.
\end{abstract}

\noindent{\bf MSC:} 46Lxx, 34B45, 35Qxx, 35R30
\smallskip

\noindent{\bf Key words:} graph of strings, controllability,
reachable sets, self-adjoint operator algebras, inverse problems
\smallskip

\noindent{\bf Short title:}  A C*-algebra associated with a graph

\setcounter{section}{-1}

\section{Introduction}
\subsubsection*{About the paper} We introduce a self-adjoint (C*-)
operator algebra algebra
$\mathfrak E$ associated with a dynamical system on a metric
graph. The system is governed by the wave equation and controlled
from the boundary vertices. Algebra $\mathfrak E$ is generated by
the so-called {\it eikonals}, which are the self-adjoint operators
related with reachable sets of the system. A structure of
reachable sets and the algebra is the main subject of the paper.
We show that $\mathfrak E$ is a direct sum of "elementary blocks".
Each block is an operator (sub)algebra, the operators multiplying
${\mathbb R}^n$-valued functions by continuous matrix-valued
functions of special kind.

The eikonal algebra is determined by the boundary dynamical and/or
spectral inverse data up to isometric isomorphism. It is an
inspiring fact, which shows promise of its possible applications
to inverse problems. In particular, one can hope to extract
information about geometry of the graph from the algebra spectrum
$\widehat{\mathfrak E}$. Such a technique works well on manifolds
\cite{BSobolev}, \cite{BJOT}, \cite{BD_2}.

The paper develops an algebraic version of the {\it boundary
control method} in inverse problems \cite{BCald},
\cite{BIP07}--\cite{BD_2}. Our approach reveals some new and
hopefully prospective relations between inverse problems on graphs
and C*-algebras.

\subsubsection*{Content}
In more detail, we deal with the
dynamical system
\begin{align*}
& u_{tt}-\Delta u=0 && {\rm in}\,\,\Omega \times (0,T)\\
&u(\,\cdot\,,t) \in {\cal K}  && {\rm
for\,\,all\,}\,t \in [0,T]\\
& u|_{t=0}=u_t|_{t=0}=0 && {\rm in}\,\, \Omega\\
& u=f && {\rm on \,}\, \Gamma \times [0,T]\,,
\end{align*}
where $\Omega$ is a finite compact metric graph, $\Gamma$ is the
set of its boundary vertices; $\Delta$ is the Laplace operator in
$\Omega$ defined on $\cal K$, which is a class of smooth functions
satisfying the Kirchhoff conditions at the interior vertices; $T
\leqslant \infty$; $f$ is a {\it boundary control}. A solution
$u=u^f(x,t)$ describes a wave initiated at $\Gamma$ by the control
$f$ and propagating into $\Omega$.

With the system one associates the {\it reachable sets} $${\cal
U}^s_\gamma\,=\,\{u^f(\,\cdot\,,s)\,|\,\,f \in L_2\left(\Gamma
\times [0,T]\right), {\rm supp\,}f \in \gamma \times [0,T]\} \quad
\gamma \in \Gamma,\, 0\leqslant s \leqslant T.$$ Let $P^s_\gamma$
be the orthogonal projection in $L_2(\Omega)$ onto ${\cal
U}^s_\gamma$. A self-adjoint operator
$$E^T_\gamma\,=\,\int_0^Ts \,dP^s_\gamma$$ is called an {\it
eikonal} (corresponding to the boundary vertex $\gamma$).

Choose a subset $\Sigma \subseteq \Gamma$. An {\it eikonal
algebra} ${\mathfrak E}^T_\Sigma$ is defined as the minimal
norm-closed C*-subalgebra of the bounded operator algebra
${\mathfrak B}\left(L_2(\Omega)\right)$, which contains all
$E^T_\gamma$ as $\gamma \in \Sigma$.
\smallskip

We provide the characteristic description of the sets ${\cal
U}^s_\gamma$ and projections $P^s_\gamma$. As a result, we clarify
how the eikonals $E^T_\gamma$ act. Thereafter, a structure of the
eikonal algebra is revealed, and we arrive at the main result: the
algebra is represented in the form of a finite direct sum
\begin{equation}\label{main result}
{\mathfrak E}^T_\Sigma\,=\,\oplus \sum_j{\mathfrak b}^T_j\,,
\end{equation}
where ${\mathfrak b}^T_j$ are the so-called {\it block algebras}.
Each ${\mathfrak b}^T_j$ is isometrically isomorphic to a
subalgebra $\tilde{\mathfrak b}^T_j \subset {\mathfrak
B}\left(L_2\left([0,\delta_j]; {\mathbb R}^{M_j}\right)\right)$
generated by the operators, which multiply elements of
$L_2\left([0,\delta_j]; {\mathbb R}^{M_j}\right)$ (vector-valued
functions of $r \in [0,\delta_j]$) by the matrix-functions of the
form $B_{\gamma,j}^* D_{\gamma,j}(r) B_{\gamma,j}\,\,\,\,(\gamma
\in \Sigma)$. Here each $B_{\gamma,j}$ is a {\it constant}
projecting matrix; $D_{\gamma,j}$ is a diagonal matrix, its
diagonal elements being the {\it linear functions} of the form
$T_{\gamma,j} \pm r$ with $T_{\gamma,j} \in (0,T]$. These
functions are continuous, and, hence, we have
\begin{equation*}\tilde{\mathfrak b}^T_j\,\subset
C\left(\left[0,\delta_j\right]; {\mathbb
M}^{M_j}\right)\,,\end{equation*} where the latter is the algebra
of continuous real $M_j \times M_j$ - matrix valued functions on
$[0, \delta_j]$.

\subsubsection*{Comments}
$\bullet$\,\,\,Algebra ${\mathfrak E}^T_\Sigma$ associated with a
graph is a straightforward analog of the eikonal algebras
associated with a Riemannian manifold: see \cite{BSobolev},
\cite{BD_2}. These algebras possess two principal features, which
enable one to apply them to solving inverse problems on manifolds:
\begin{enumerate}
\item the eikonal algebra is determined (up to isometric
isomorphism) by dy\-na\-mi\-cal and/or spectral boundary inverse
data \item its spectrum is, roughly speaking, identical to the
manifold.
\end{enumerate}
By this, one can solve the problem of reconstruction of the
manifold via its inverse data by the scheme
\cite{BIP07}--\cite{BD_2}: $${\rm data}\,\Rightarrow\,\,{\rm
relevant\,\, eikonal\,\, algebra}\, {\mathfrak E}\,
\Rightarrow\,{\rm its\,\, spectrum}\, \widehat {\mathfrak
E}\,\,\equiv {\rm manifold}.$$ It is so effective application,
which has motivated to extend this approach to inverse problems on
graphs. The hope was that a graph seems to be a simpler object
than a manifold of arbitrary dimension and topology.

Surprisingly, the latter turns out to be an illusion. First of
all, in contrast to the eikonal algebras on
manifolds\footnote{somehow or other, these algebras are reduced to
the algebra $C(\Omega)$ of continuous functions}, the algebra
${\mathfrak E}^T_\Sigma$ is {\it noncommutative}. By this, in the
general case, its spectrum $\widehat{\mathfrak E}^T_\Sigma$
endowed with the Jacobson topology is a {\it non-Haussdorff}
space. Hence, $\widehat{\mathfrak E}^T_\Sigma$ is by no means
identical to the (metric) graph $\Omega$, so that property 2
fails.

However, property 1 does hold. Also, the known examples show that
representation (\ref{main result}) and structure of the spectrum
$\widehat{\mathfrak E}^T_\Sigma$ reflect some features of the
graph geometry. Therefore, an attempt to extract information on
$\Omega$ from the eikonal algebra and, eventually, to recover
$\Omega$ seems quite reasonable. Hopefully, our paper is a step
towards this goal.
\medskip

\noindent$\bullet$\,\,\,In view of big volume of the paper, we
omit the proofs of some technical propositions.
\medskip

\noindent$\bullet$\,\,\,This work is supported by grants RFBR
11-01-00407A and SPbGU\\ 11.38.63.2012, 6.38.670.2013. We are
grateful to the University of Kyoto and personally Prof. Yu.Iso
for kind support of our collaboration. We thank I.V.Kubyshkin for
the help in computer graphics.

\section{Graph}
\subsection{Basic definitions}
\subsubsection{Standard star}
Let $I_j:=(0,a_j)=\{s \in {\mathbb R}\,|\,\,0<s<a_j<
\infty\},\,\,j=1,\,\dots\,,m$ be finite intervals, each interval
being regarded as a subspace of the metric space $\mathbb R$ with
the distance $|s-s'|$. The set $S_m:=\{0\}\cup I_1 \cup \dots \cup
I_m$ endowed with the metric
\begin{equation*}
{\rm dist}\,(s,s')\,:=\,
\begin{cases}
|s-s'| \qquad &  s,s' \in I_j\\
s+s' \qquad & s \in I_i,\, s' \in I_j,\,\,\,i \not= j\\
s'\qquad & s=0,\, s' \in I_j\\ s \qquad & s \in I_i,\,\,s'=0\\0
\qquad & s=s'=0
\end{cases}
\end{equation*}
is called  a (standard) {\it $m$-star} (see Fig.1a). Note that a
2-star is evidently isometric to the interval $(-a_1, a_2)$.

\begin{figure}
 \begin{center}
 \epsfysize=8cm
 \epsfbox{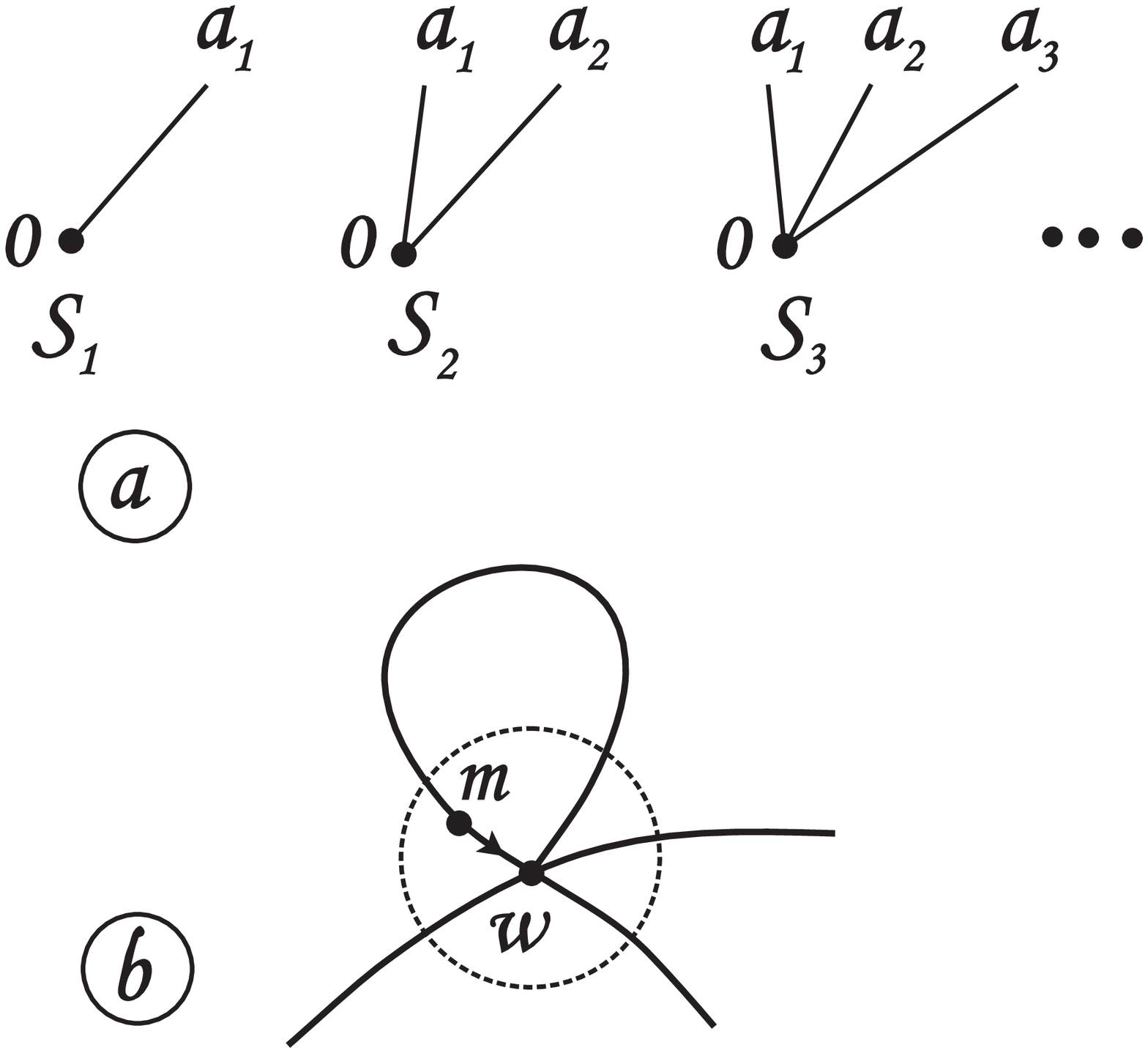}
 \end{center}
\caption{Stars and star neighborhoods}
 \end{figure}

\subsubsection{Metric graph}
A compact connected metric space $\Omega$ with the metric $\tau:
\Omega \times \Omega \to [0,\infty)$ is said to be a {\it
homogeneous metric graph} if the following is fulfilled:
\begin{itemize}
\item $\Omega =E \cup V\cup \Gamma $, where $E=\left \{e_j \right
\}_{j=1}^p$, $e_j$ are the {\it edges}; $V=\left \{v_k \right
\}_{k=1}^q$, $v_k$ are the {\it interior vertices}; $\Gamma =\left
\{ \gamma _l \right \}_{l=1}^n$, $\gamma_l$ are the {\it boundary
vertices}, $\Gamma \not=\emptyset$ \item each $e_j$ is isometric
to a finite interval $\{s \in {\mathbb R}\,|\,\,a_j < s < b_j\}$
\begin{Convention}\label{Conv 1}
In what follows, we assume that the isometries ({\it
para\-met\-ri\-za\-tions}) $\eta_j:\, e_j \to (a_j,b_j)$ are fixed
and write ${\tilde y}(s):=(y\circ \eta^{-1}_j)(s),\,\,\,s \in
(a_j,b_j)$ for a function $y=y(x)$ on $\Omega$ restricted to the
edge $e_j$.
\end{Convention}
So, for $x,x' \in e_j$ one has
$\tau(x,x')=|\eta_j(x)-\eta_j(x')|$. \item every $w \in
V\cup\Gamma$ has a neighborhood (in $\Omega$) isometric to an
$S_m$ with $m \geqslant 3$ or $m=1$, the number $m=m(w)$ being
called a {\it multiplicity} of $w$. The vertices of multiplicity
$\geqslant 3$ constitute the set $V$. The vertices of multiplicity
$1$ form the {\it boundary} $\Gamma$.

As was noted above, 2-stars are isometric to intervals (edges). By
this, they do not take part in further considerations.
\end{itemize}
For a point $x\in \Omega$, by $\Omega^r[x]:=\{x' \in
\Omega\,|\,\tau(x,x')<r\}$ we denote its metric neighborhood of
radius $r>0$. For a subset $A \subset \Omega$, we put
$\Omega^r[A]:=\{x \in \Omega\,|\,\tau(x,A)<r\}$.

Let the points $x \not=x^\prime$ belong to a (parametrized) edge
$e$. The set $\eta^{-1}\left(\left(\min\{\eta(x),
\eta(x^\prime)\}, \max\{\eta(x),
\eta(x^\prime)\}\right)\right)\subset e$ is called an {\it
interval} and denoted by $]x, x^\prime[$.

\subsubsection{Characteristic set}
In the space-time $\Omega \times {\overline {\mathbb R}}_{+} $,
for a fixed $(x_0,t_0)$ define a {\it characteristic cone}
$$
{\rm ch\,}[(x_0,t_0)] :=\left\{(x,t) \, | \,\, t-t_0 =\tau
(x,x_0)\right\}\,;
$$
for a subset $A \subset \Omega \times {\overline {\mathbb
R}}_{+}\,\,\,$ put
$$
{\rm ch\,}[A]\,:= \,
       \bigcup\limits_{(x,t)\, \in \, A} {\rm ch\,}[(x,t)]\,.
$$
A {\it characteristic set} ${\rm Ch\,}[(x_0,t_0)]$ is introduced
by the following recurrent procedure:

\noindent{\bf Step 0:}\,\,put
$$
C^0[(x_0,t_0)]\,:=\, {\rm ch\,}[(x_0,t_0)]
$$
and
$$
W^0(x_0,t_0)\,:=\,
          \left\{(w,t) \in C^0[(x_0,t_0)]\,|\,\,w \in V\cup\Gamma \right\}\,;
$$

\noindent{\bf Steps j\,=\,1,2, $\dots$:}\,\,put
$$
C^j[(x_0,t_0)]\,:=\,
      C^{j-1}[(x_0,t_0)] \bigcup
           {\rm ch\,}\left[W^{j-1}(x_0, t_0)\right]
$$
and
$$
W^j(x_0,t_0)\,:=\,
          \left\{(w,t) \in C^{j-1}[(x_0,t_0)]\,|\,\,w \in V\cup\Gamma \right\}\,;
$$
$$
.......................................\,\,\,.
$$
At last, define
$$
{\rm Ch\,}[(x_0,t_0)]\,:=\,\bigcup\limits_{j=0}^\infty
               C^j[(x_0,t_0)]\,\,\,\,.
$$

Note that ${\rm Ch\,}[(x_0,t_0)]$ can be also characterized as the
minimal subset in $\Omega \times {\overline {\mathbb R}}_{+}$
satisfying the conditions:
\begin{itemize}
\item ${\rm ch\,}[(x_0,t_0)] \subset {\rm Ch\,}[(x_0,t_0)]$ \item
if $w\in V\cup\Gamma$ and $t_w \in {\mathbb R}_{+}$ are such that
$(w,t_w)\in {\rm Ch\,}[(x_0,t_0)]$ then ${\rm
ch\,}[(w,t_w)]\subset {\rm Ch\,}[(x_0,t_0)]$. \end{itemize} The
characteristic set can be regarded as a space-time graph; Fig.2
illustrates the case $x_0=\gamma \,,\,\, t_0=0$.

\begin{figure}
 \begin{center}
 \epsfysize=8cm
 \epsfbox{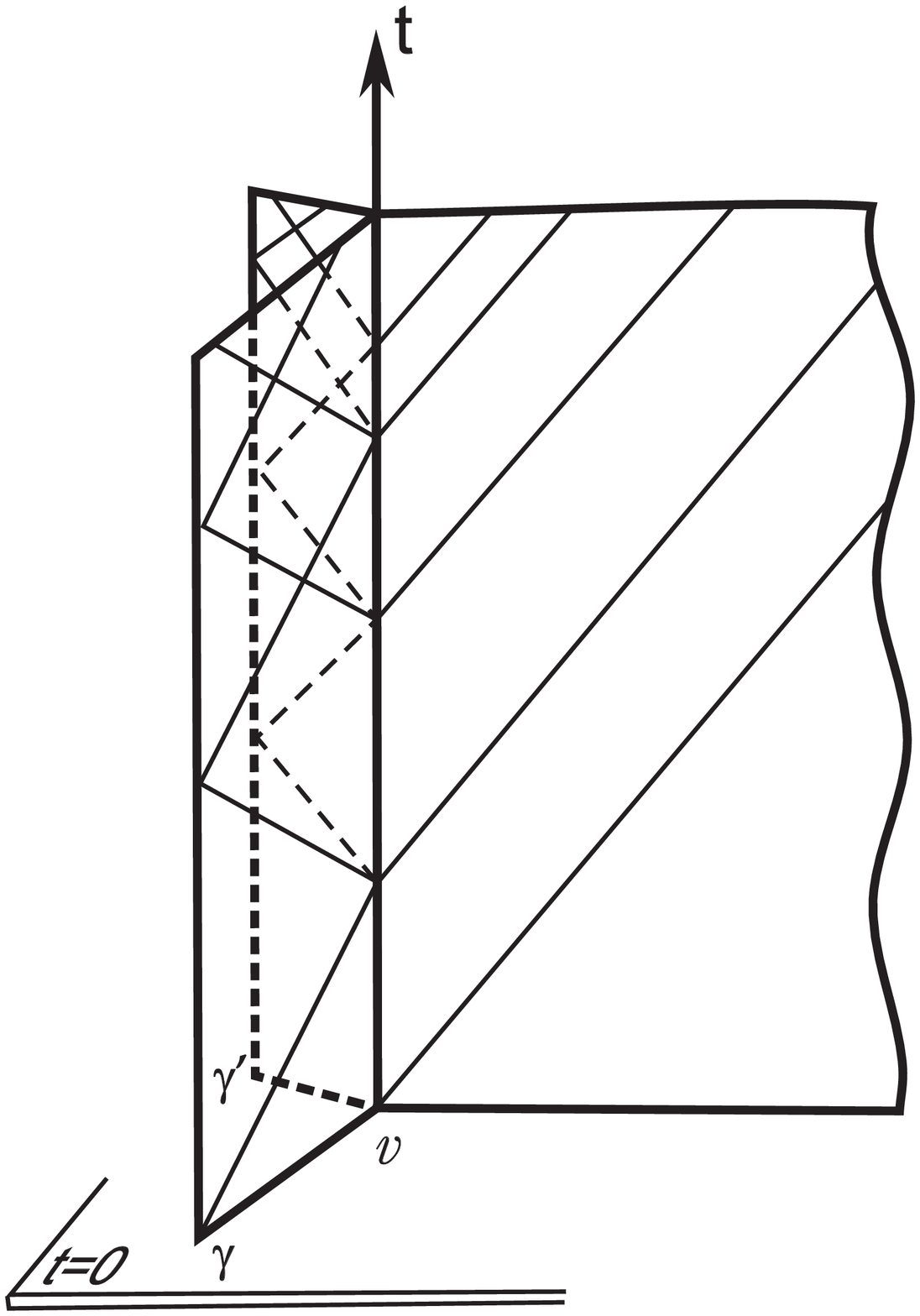}
 \end{center}
\caption{Characteristic set}
 \end{figure}

Such a graph is also a metric space: it is endowed with the length
element
\begin{equation}\label{metric on Char Set}
d\nu^2\,:=\,d\tau^2\,+\,dt^2\,,
\end{equation}
i.e., for the close points $(x,t), (x',t')\in {\rm
Ch\,}[(x_0,t_0)]$ one has $\nu\left((x,t),
(x',t')\right)=\left[\tau^2(x, x')+(t-t')^2\right]^{1\over 2}$.
For arbitrary points, the distance $\nu$ is defined as the length
of the shortest curves lying in ${\rm Ch\,}[(x_0,t_0)]$ and
connecting the points.

\subsection{Spaces and operators}
\subsubsection{Derivatives}
For an edge $e \in E$ parametrized by $\eta: e \to (a,b)$, a
function $y$ on $\Omega$, and a point $x \in e$, we define
\begin{equation*}
\frac{dy}{de}(x)\,:=\,\frac{d\tilde
y}{ds}\bigg|_{s=\eta(x)}\,=\,\lim \limits_
       {x' \to x \atop
       \eta(x')>\eta(x)}
       \frac{y(x')-y(x)}{\tau(x',x)}
\end{equation*}
(recall that ${\tilde y}:=y \circ \eta^{-1}$).
\medskip

Fix a vertex $w \in V \cup \Gamma$  and choose its neighborhood
$\omega \subset \Omega$ isometric to $S_m$. We say an edge $e$ to
be {\it incident} to $w$ if ${\overline e} \ni w$ or,
equivalently, if $e \cap \omega \not=\emptyset$. Note that $e \cap
\omega$ can consist of two components (see Fig.1b) and settle
that, in this case, each component is regarded as a single edge
(of the subgraph $\omega$) incident to $w$.

For every $e$ incident to $w$, define an {\it outward derivative}
\begin{equation*}
\frac{dy}{de_+}(w)\,:=\,\lim \limits_
       {e \ni m \to w} \frac{y(m)-y(w)}{\tau(m,w)}\,.
\end{equation*}
For an interior vertex $v \in V$ and a function $y$, define an
{\it outward flow}
\begin{equation*}
\Pi_v[y]\,:=\, \sum \limits_{{\overline e}\, \ni v}
\frac{dy}{de_+}(v)\,,
\end{equation*}
the sum being taken over all edges incident to $v$ in a star
neighborhood $\omega \ni v$.

\subsubsection{Spaces}
Introduce a (real) Hilbert space ${\cal H}:=L_2(\Omega)$ of
functions on $\Omega$ with the inner product
$$
(y,u)_{\cal H}=\int _\Omega y u\,d\tau = \sum_{e \in E}\,\,\int _e
y u\,d\tau := \sum_{e \in E}\,\,\int \limits_{\eta(e)} {\tilde
y}(s)\,{\tilde u}(s)\,ds\,.
$$
By $C(\Omega)\subset \cal H$ we denote the class of functions
continuous on $\Omega$.

We assign a function $y$ on $\Omega $ to a class ${\cal H}^2$ if
$y \in C(\Omega)$ and ${\tilde y}|_{\eta(e)} \in
H^2({\eta(e)})$\footnote{$H^s(a,b)$ is the standard Sobolev
space.} for each $e \in E$.

Also, define the {\it Kirchhoff class}
\begin{align}\label{1.4}
{\cal K} \,:= \,\{y \in {\cal H}^2\,|\,\,\Pi_v[y]=0,\,\,\,v \in V
\}\,.
\end{align}

\subsubsection{Operator}
The {\it Laplace operator} on the graph
 $\Delta: {\cal  H}\to{\cal  H},
\,\, {\rm Dom}\, \Delta={\cal  K}$,
\begin{equation}\label{Delta=d2/de2}
\left (\Delta y\right )\big|_e\,:=\,\frac{d^2y}{de^2}, \qquad e
\in E\,
\end{equation}
is well defined (does not depend on the parametrizations). It is a
closed densely defined operator in $\cal H$.

\section{Waves on graph}
\subsection{Dynamical system} An initial boundary value
problem of the form
\begin{align}\label{Graph 1}
& u_{tt}-\Delta u=0 && {\rm in}\,\,\left[\Omega \setminus (V\cup
\Gamma)\right] \times (0,T)\\
\label{Graph 2} &u(\,\cdot\,,t) \in {\cal K}  && {\rm
for\,\,all\,}\,t \in [0,T]\\
\label{Graph 3} & u|_{t=0}=u_t|_{t=0}=0 && {\rm in}\,\, \Omega\\
\label{Graph 4} & u=f && {\rm on \,}\, \Gamma \times [0,T]
\end{align}
is referred to as a dynamical system associated with the graph
$\Omega$. Here $T < \infty$; $f=f(\gamma ,t)$ is a {\it boundary
control}; the solution $u=u^f(x,t)$ describes a {\it wave}
initiated at $\Gamma$ and propagating into $\Omega$.

Note that, by definition (\ref{1.4}), the condition (\ref{Graph
2}) provides the {\it Kirchhoff laws}:
\begin{equation}\label{Kirchhoff laws}
u(\,\cdot\,,t) \in C(\Omega), \qquad \Pi_v[u(\,\cdot\,,t)]=0
\end{equation}
for all $t \geqslant 0$ and $v \in V$. Also, note that by
(\ref{Delta=d2/de2}), on each edge $e \in E$ parametrized by
$\eta: e \to (a,b)$, the pull-back function $\tilde u
(\,\cdot\,,t)=u (\,\cdot\,,t)\circ\eta^{-1}$ satisfies the
homogeneous string equation
\begin{equation}\label{string eqn}
\tilde u_{tt}-\tilde u_{ss}=0 \qquad {\rm
in\,}\,\,(a,b)\times (0,T).
\end{equation}
It is the reason, by which we regard $\Omega$ as a graph
consisting of {\it homogeneous strings}. As is well known, for a
$C^2$-smooth (with respect to $t$) control $f$ vanishing near
$t=0$ the problem has a unique classical solution $u^f$.

\smallskip
A space of controls ${\cal F}^T :=L_2\, (\Gamma \times [0,T])$
with the inner product
$$
(f,g)_{{\cal F}^T} :=\sum\limits_{\gamma \in \Gamma} \,
\int\limits_{0}^T f(\gamma ,t)\,g(\gamma ,t)\,dt
$$
is called an {\it outer space} of system (\ref{Graph
1})--(\ref{Graph 4}). It contains the subspaces ${\cal
F}^T_\gamma:=\left\{f \in {\cal F}^T\,|\,\, {\rm supp\,}f \subset
\{\gamma\} \times [0,T]\right\}$ of controls, which act from
single boundary vertices $\gamma \in \Gamma$, so that
\begin{equation}\label{F=sum F gamma}
{\cal F}^T\,=\,\oplus \sum_{\gamma \in \Gamma}{\cal F}^T_\gamma
\end{equation}
holds. Each $f \in {\cal F}^T_\gamma$ is of the form
$f(\gamma',t)=\delta_\gamma(\gamma')\varphi(t)$ with $\varphi \in
L_2(0,T)$.
\smallskip

The space ${\cal H}$ is an {\it inner space}; the waves
$u^f(\,\cdot\, ,t)$ are time--dependent elements of ${\cal H}$.

\subsection{Fundamental solution}
\subsubsection{Definition}
Consider the system (\ref{Graph 1})--(\ref{Graph 4}) with
$T=\infty$.

For $\gamma, \gamma' \in \Gamma$, we denote
\begin{equation*} \delta_\gamma (\gamma'):= \begin{cases} 0, &
\gamma'\not= \gamma
\\1, & \gamma'=\gamma\end{cases} \, ; \end{equation*} let
$\delta(t)$ be the Dirac delta-function of time.

Fix a boundary vertex $\gamma$. Taking the (generalized) control
$f(\gamma',t)=\delta_\gamma (\gamma')\delta(t)$, one can define
the generalized solution $u^{\delta_\gamma \delta}$ to (\ref{Graph
1})--(\ref{Graph 4}). A possible way is to use a smooth
regularization $\delta^\varepsilon(t) \underset{\varepsilon \to
0}\to \delta(t)$ and then understand $u^{\delta_\gamma \delta}$ as
a relevant limit of the classical solutions $u^{\delta_\gamma
\delta^\varepsilon}$ as $\varepsilon \to 0$. Such a limit turns
out to be a space-time distribution on $\Omega \times [0,T]$ of
the class $C\left((0,T); H^{-1}(\Omega)\right)$ (see, e.g.,
\cite{ContrGraph}).

The distribution $u^{\delta_\gamma \delta}$ is called a {\it
fundamental} solution to (\ref{Graph 1})--(\ref{Graph 4})
(corresponding to the given $\gamma$). It describes the wave
initiated by an instantaneous source supported at $\gamma$.
Consider its properties in more detail; all of them are well
known.  Recall that $\tau$ is the distance in $\Omega$.

\subsubsection{First edge}
Let $e$ be the edge incident to $\gamma$ and parametrised by
$s=\eta(x):=\tau(x, \gamma)\in (0,\tau(\gamma,v))$. Let $ v\in V$
be the second vertex incident to $e$. For times $0<t\leqslant \tau
(\gamma ,v)$, by (\ref{string eqn}) one has
\begin{align*}
& \tilde u_{tt}-\tilde u_{ss}=0 && {\rm in\,}\,\,(0,\tau (\gamma
,v))\times (0,T)\\
& \tilde u|_{t=0}=\tilde u_t|_{t=0}=0 && {\rm in\,}\,\,[0,\tau
(\gamma
,v)]\\
& \tilde u|_{s=0}= \delta(t), && 0 \leqslant t \leqslant \tau
(\gamma ,v),
\end{align*}
which implies $\tilde u(s,t)=\delta(t-s)$. This evidently leads to
the representation
\begin{equation}\label{first edge}
u^{\delta_\gamma
\delta}(\,\cdot\,,t)\,=\,\delta_{x(t)}(\,\cdot\,)\,, \qquad 0
\leqslant t \leqslant \tau (\gamma ,v)\,,
\end{equation}
where $x(t)$ belongs to $e$ and satisfies $\tau\left(x(t),
\gamma\right)=t$, $\delta_p \in H^{-1}(\Omega)$ is the Dirac
measure supported at $p \in \Omega$. It means that the
$\delta$-singularity, which is injected into the graph from
$\gamma$, moves along $e$ towards $v$ with velocity $1$ (see
Fig.3a).

\begin{figure}
 \begin{center}
 \epsfysize=8cm
 \epsfbox{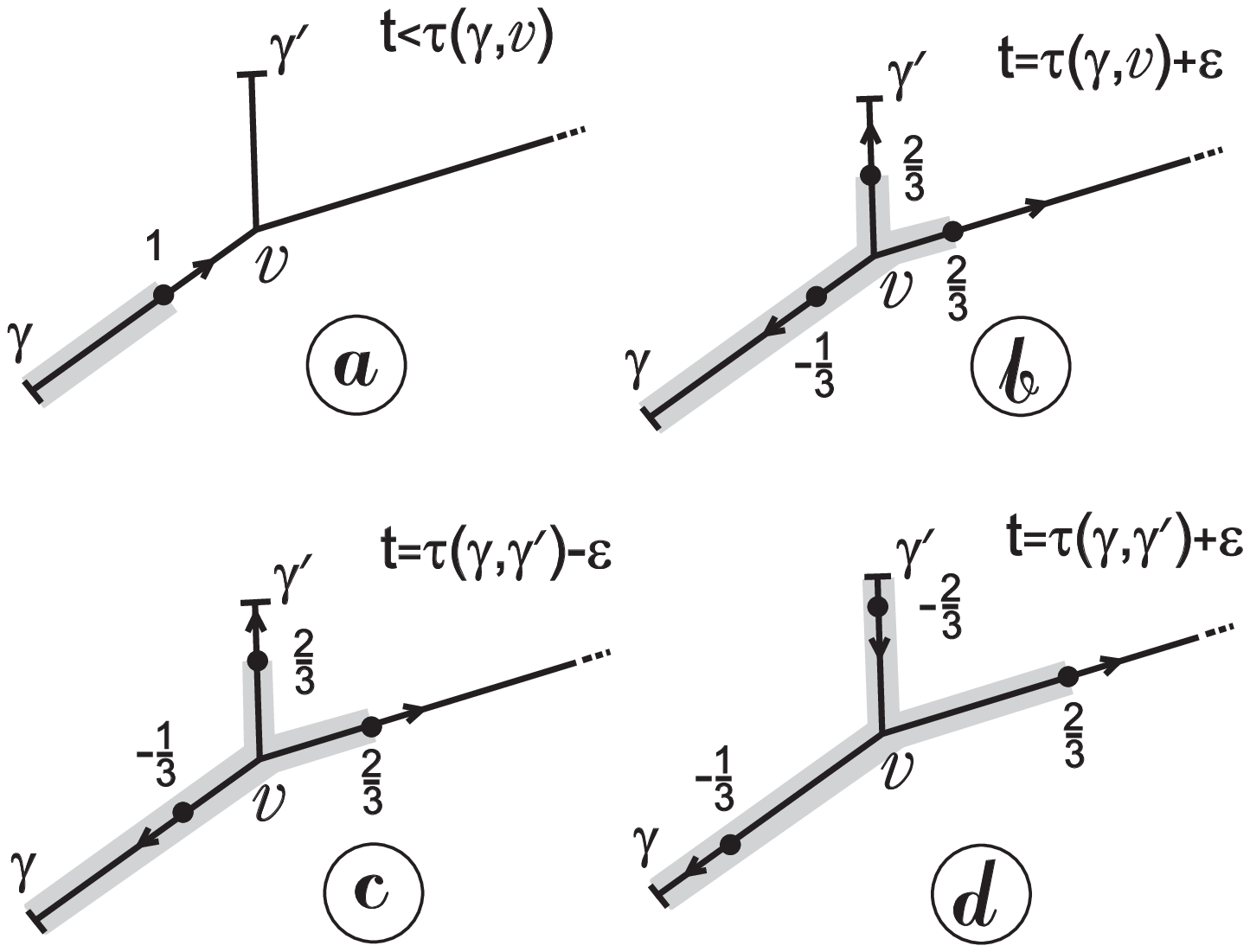}
 \end{center}
\caption{Propagation of singularities}
 \end{figure}

\subsubsection{Passing through interior vertex}
At the moment $t=\tau (\gamma ,v)$ the singularity reaches $v$ and
then passes through $v$. A simple analysis using (\ref{Kirchhoff
laws}) and (\ref{string eqn}) provides
\begin{equation}\label{Passing v}
u^{\delta_\gamma \delta}(\,\cdot\,,
t)=\sum\limits_{e':\,\overline{e'}\ni
v}a(x_{e'}(t))\,\delta_{x_{e'}(t)}(\,\cdot\,)\,,  \quad  \tau
(\gamma ,v) <t\leqslant \tau (\gamma ,v)+\varepsilon\,,
\end{equation}
as $\varepsilon >0$ is small (namely, $\varepsilon < \tau(v,(V\cup
\Gamma)\backslash\{v\})$), where $x_{e'}(t)$ belongs to $e'$ and
satisfies $\tau(x_{e'}(t),v)=t-\tau(\gamma,v)$. The function ({\it
amplitude}) $a$ is
\begin{equation}\label{amplitude a}
a(x)=\begin{cases} -\frac{m(v)-2}{m(v)} & {\rm as}\,\,\,x \in e\\
\frac{2}{m(v)} & {\rm as}\,\,\,x \in e':\,\,e'\not=e,\,\,\,
\overline{e'}\ni v
\end{cases}.
\end{equation}
Hence, in passing through $v$, the singularity splits onto $m(v)$
parts (singularities), the first one is reflected back into $e$,
the others are injected into the other $m(v)-1$ edges $e'$
incident to $v$. The reflected singularity has the negative
amplitude. The process is illustrated by Fig.3b. Note that the
`conservation law'
\begin{equation}\label{conservation law}-\frac{m(v)-2}{m(v)}+
(m(v)-1)\,\frac{2}{m(v)}\,=\,1,
\end{equation}
is valid, so that the total amplitude after the passage through
$v$ is equal to the amplitude of the incident singularity.

In what follows, we refer to (\ref{Passing v})--(\ref{conservation
law}) as a {\it splitting rule}.

\subsubsection{Reflection from boundary}
Let $\gamma' \in \Gamma$ be a boundary vertex nearest to $v$:
$$
\tau (\gamma',v )= \min\limits_{\gamma''\in \Gamma} \,\tau
(\gamma'',v)
$$
(may be $\gamma=\gamma'$), so that $\tau (\gamma ,\gamma')=\tau
(\gamma ,v)+\tau (v,\gamma')$ holds. Let $e'$ be the edge incident
to $\gamma'$.

As $t\to \tau (\gamma ,\gamma')-0$, one of the singularities,
which have appeared as a result of passing through $v$ (and,
perhaps, through another vertices or reflected from $v$ back to
$\gamma$), approaches to $\gamma'$ (see Fig.3c). Then this
singularity is reflected from $\gamma'$. A simple analysis with
the use of the condition $u^{\delta_\gamma \delta}
(\gamma',t)=\delta_\gamma(\gamma')\delta(t)=0,\,\,\,t>0$ leads to
the representation
\begin{equation}\label{Reflection from Gamma}
u^{\delta_\gamma \delta}(\,\cdot\,,t)\,=\,\begin{cases}
a\,\delta_{x(t)}(\,\cdot\,) & {\rm as}\,\,
t \in \left(\tau (\gamma, \gamma')-\varepsilon, \tau (\gamma,\gamma')\right)\\
- a\,\delta_{x(t)}(\,\cdot\,) & {\rm as}\,\, t \in \left(\tau
(\gamma,\gamma'), \tau (\gamma, \gamma')+\varepsilon\right)\,,
  \end{cases}
\end{equation}
where $x(t) \in e'$ satisfies
$\tau\left(x(t),\gamma'\right)=|t-\tau
(\gamma,\gamma')|,\,\,\,a={\rm const\,}\not=0$.

Thus, as a result of reflection from a boundary vertex, the
singularity moves {\it from} it and changes the sign of the
amplitude (see Fig.3d). This is a {\it reflection rule}.
\smallskip

The splitting and reflection rules, along with superposition
principle (linearity of the system), uniquely determine the
fundamental solution $u^{\delta_\gamma \delta}$ for all $t
\geqslant 0$. Let us list some of its well-known properties.

\subsubsection{Hydra}
Return to the fundamental solution of (\ref{Graph 1})--(\ref{Graph
4}) with $T=\infty$ and consider $u^{\delta_\gamma \delta}$  as a
space-time distribution in $\Omega \times \overline{{\mathbb
R}_+}$. In what follows, its support
$$H_\gamma\,:=\,{\rm supp\,}u^{\delta_\gamma \delta}$$ plays
important role and is called a {\it hydra}, the point $(\gamma,
0)$ being its {\it root}. The reason to introduce the
characteristic set (see 1.1.3) is that it consists of the
characteristic lines of the wave equation (\ref{Graph 1}). As is
seen from (\ref{first edge})--(\ref{Reflection from Gamma}),
singularities propagate along the characteristics that leads to
the relation
\begin{equation*}\label{supp Hydra}
H_\gamma\,\subseteq \,{\rm Ch\,}[(\gamma, 0)]\,.
\end{equation*}
In particular, it shows that the space projections of
singularities propagate along a homogeneous graph with velocity
$1$. This implies
\begin{equation}\label{supp u delta delta}
{\rm supp\,}u^{\delta_\gamma \delta}(\,\cdot\,,t)\, \subset
\overline{\Omega^t[\gamma]}\,, \qquad t>0\,,
\end{equation}
where the left hand side is understood as a time-depended element
of $H^{-1}(\Omega)$. Also, note that the hydra is a connected set
in $\Omega \times [0,T]$: as is evident, any point $(x,t) \in
H_\gamma$ is connected with the root $(\gamma,0)$ through a path
in $H_\gamma$.

Let $$\pi: H_\gamma \to \Omega,\,\, \pi\left((x,t)\right):=x\,,
\qquad \rho: H_\gamma \to \overline{{\mathbb R}_+},\,\,
\rho\left((x,t)\right):=t$$ be the {\it space} and {\it time
projection} respectively. For $A \subset \Omega$ and $B \subset
\overline{{\mathbb R}_+}$, denote
$$\pi^{-1}(A)\,:=\,\left\{(x,t)\in H_\gamma\,|\,\,x \in A\right\}, \,\,\,
\rho^{-1}(B)\,:=\,\left\{(x,t)\in H_\gamma\,|\,\,t \in B\right\}
.$$

Choose an edge $e \in E$ parametrized by $\eta: e \to (a,b)$. Its
pre-image $\pi^{-1}(e)\subset H_\gamma$ consists of the sets
$$\tilde e_j\,:=\,\{\left(\eta^{-1}(s),t_j + \sigma_j\,
\left(s-a\right)\right)\,|\,\,a<s<b\}\,, \qquad \sigma_j=\pm 1,
\quad j=1,2, \dots\,,$$ which are the edges of the hydra as a
space-time graph. There is a part of the fundamental solution of
the form
\begin{equation}\label{FSol on e}
u^{\delta_\gamma \delta}_j(\,\cdot\,,t)= a_j\,
\delta_{x(t)}(\,\cdot\,)\,, \qquad t \in \rho (\tilde e_j)
\end{equation} supported on $\tilde e_j$, where $a_j={\rm
const}\not=0$, $x(t):=\eta^{-1}\left(a+\sigma_j(t-t_j)\right)$. In
dynamics, (\ref{FSol on e}) describes the singularity moving along
$e$ with velocity $1$ as $t$ runs over the time interval $\rho
(\tilde e_j)$, the sign $\sigma_j$ determining the direction of
motion. The value of the amplitude $a_j$ is determined by the
prehistory of $u^{\delta_\gamma \delta}$ as $t <\inf \rho (\tilde
e_j)$ and can be derived from the splitting and reflection rules.
\smallskip

\noindent{\bf Amplitude.}\,\,\,The aforesaid (see (\ref{first
edge}), (\ref{Passing v}), (\ref{Reflection from Gamma}),
(\ref{FSol on e})) enables one to endow the hydra with a function
$a$ ({\it amplitude}) as follows. Take a point $(x,t)\in H_\gamma$
provided $x \in \Omega \backslash \{V \cup \Gamma\}$, so that
there is a graph edge $e \ni x$. Hence, there is a hydra edge
$\tilde e_j \subset \pi^{-1}(e)$ such that $(x,t)\in \tilde e_j$
and (\ref{FSol on e}) does hold. In the mean time, as is easy to
see, there may be at most one more edge $\tilde e_i \in
\pi^{-1}(e),\,\,\tilde e_i \not= \tilde e_j$, which contains the
given $(x,t)$ (so that $(x,t)= \tilde e_i \cap \tilde e_j$: see
the point $p$ on Fig.4). Then,
\begin{itemize} \item ({\it generic case})\,\,\,if $\tilde e_j$ is
a unique edge, which contains $(x,t)$, we set $a(x,t):=a_j$ \item
({\it exclusive case})\,\,\,if there is the second $\tilde e_i \ni
(x,t)$, we define $a(x,t):=a_i+a_j$
\end{itemize}
(on Fig.4, $a(p)=-\frac{4}{9}+\frac{1}{3}=-\frac{1}{3}$). To
extend the amplitude to the whole hydra we add the following:
\begin{itemize} \item  for points $(\gamma', t) \in H_\gamma$
with $\gamma' \in \Gamma$, we put $a(\gamma',t):=0$ as $t>0$, and
$a(\gamma',0):=\delta_\gamma(\gamma')$ (Kronecker's symbol) \item
if $(v,t) \in H_\gamma$ and $v \in V$, we define
$a(v,t)=a_1+\dots+a_p$, where $a_i$ are the amplitudes (\ref{FSol
on e}) on the hydra edges $\tilde e_i \subset H_\gamma \cap
\{(x',t')\,|\,\,t'<t\}$ incident to $(v,t)$. Note that in the
generic case such an $\tilde e_i$ is unique.

\end{itemize}
\begin{figure}
 \begin{center}
 \epsfysize=7cm
\epsfxsize=6cm
 \epsfbox{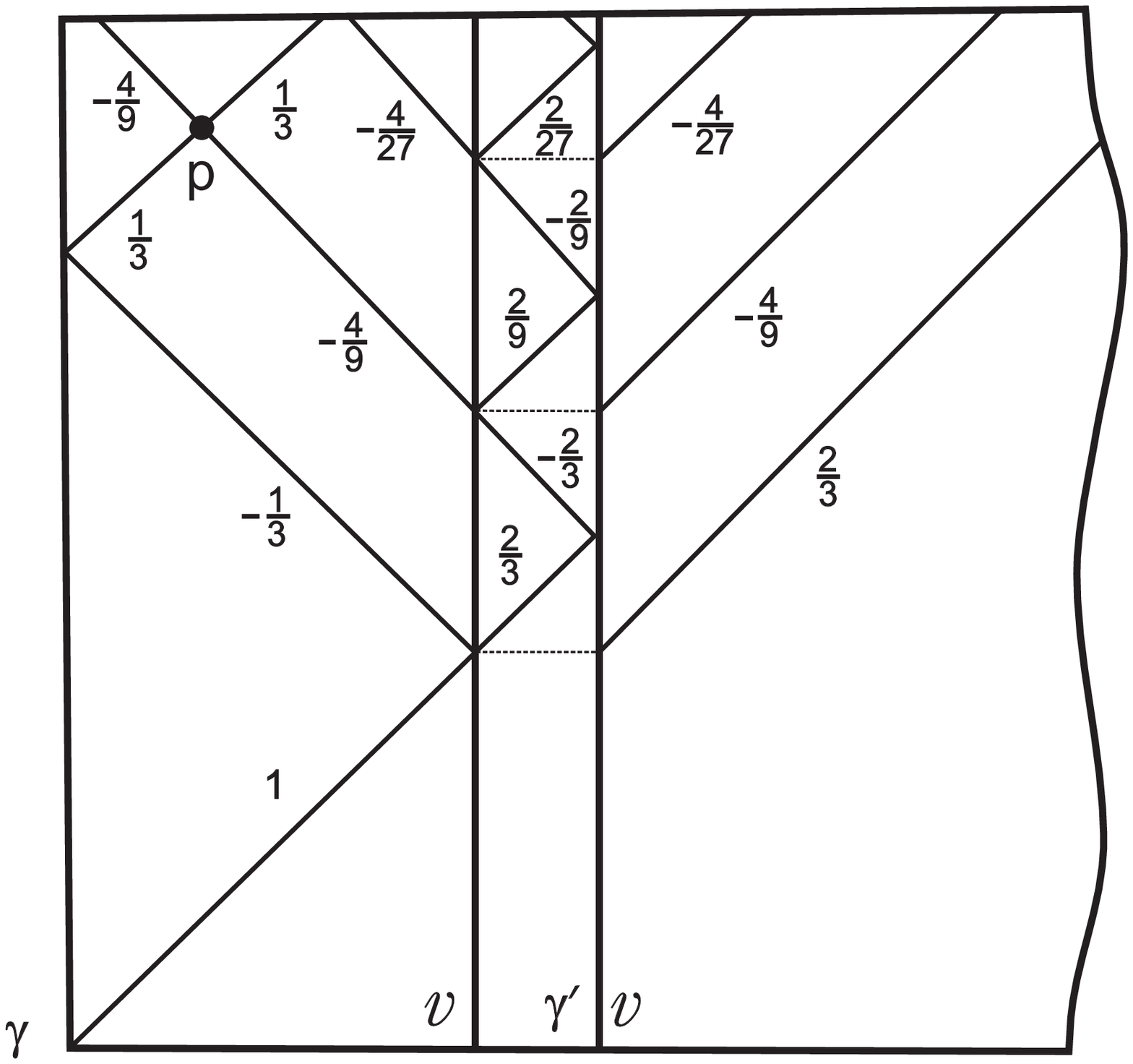}
 \end{center}
\caption{Hydra and amplitude function}
 \end{figure}

\noindent{\bf Corner points.}\,\,\, Thus, the amplitude $a$ is a
well-defined piece-wise constant function on $H_\gamma$. Moreover,
as a function on the metric space-time graph (see (\ref{metric on
Char Set})), it is piece-wise continuous, the continuity being
broken only in some exceptional points. Namely, we say $(x,t) \in
H_\gamma$ to be a {\it corner point} if either $x \in \Gamma \cup
V$ or there are an edge $e \ni x$ and the hydra edges $\tilde e_i,
\tilde e_j \subset \pi^{-1}(e)$ such that $(x,t)=\tilde e_i \cap
\tilde e_j$ (see Fig.4).

\subsection{Generalized solutions}
Here we list some results on solutions of the problem (\ref{Graph
1})--(\ref{Graph 4}), which can be easily derived from the above
mentioned properties of the fundamental solution. Unless otherwise
specified, we deal with $T<\infty$.
\smallskip

\subsubsection{Definition}
Fix a $\gamma \in \Gamma$. Let a control $f \in {\cal F}^T_\gamma$
(see (\ref{F=sum F gamma})) be such that $u^f$ is a classical
solution. Then the {\it Duhamel representation}
\begin{equation}\label{Duhamel}
             u^f\, = \,u^{\delta_\gamma \delta}
             \underset{t}\ast f, \qquad {\rm in\,}\,\,
             \Omega \times [0,T]
\end{equation}
(the convolution with respect to time) holds and motivates the
following. For an $f \in {\cal F}^T$, we {\it define} a
(generalized) solution to (\ref{Graph 1})--(\ref{Graph 4}) by
\begin{equation}\label{Duhamel General}
             u^f\, := \,\sum \limits_{\gamma \in \Gamma} u^{\delta_\gamma \delta}
             \underset{t}\ast f, \qquad {\rm in\,}\,\,
             \Omega \times [0,T]\,.
\end{equation}

\subsubsection{General properties}
\begin{enumerate}
\item As can be shown, solution (\ref{Duhamel General}) belongs to
the class $C\left([0,T]; {\cal H}\right)$, i.e., is a continuous
$\cal H$-valued function of time. \item For $f \in {\cal
F}^T_\gamma$, relation (\ref{supp u delta delta}) implies
\begin{equation}\label{supp u^delta delta}
{\rm supp\,}u^f(\,\cdot\,,t)\, \subset
\overline{\Omega^t[\gamma]}\,, \qquad t>0\,,
\end{equation}
which means that the waves propagate in $\Omega$ with velocity
$1$. Let $\Sigma \subseteq \Gamma$ be a set of boundary vertices
and $f \in \oplus \sum_{\gamma \in \Sigma}{\cal F}^T_\gamma$. As a
consequence of (\ref{supp u^delta delta}), we have the relation
\begin{equation}\label{supp u^f}
{\rm supp\,}u^f(\,\cdot\,,t)\, \subset \bigcup \limits_{\gamma \in
\Sigma} \overline{\Omega^t[\gamma]}=\overline{\Omega^t[\Sigma]}\,,
\qquad t>0\,,
\end{equation}
which is usually referred to as a {\it finiteness of domain of
influence}. \item For the rest of the paper we accept the
following.
\begin{Convention}\label{Conv 2}
All functions depending on time $t \geqslant 0$ are extended to
$t<0$ by zero.
\end{Convention}
For $f \in {\cal F}^T$, denote by $f_s(\gamma,t):=f(\gamma, t-s)$
the delayed control. Since the graph and operator $\Delta$, which
governs the evolution of system (\ref{Graph 1})--(\ref{Graph 4}),
do not depend on time, one has the relation ({\it steady-state
property})
\begin{equation}\label{steady state}
 u^{f_s}(\,\cdot\,,t) = u^f(\,\cdot\,,t-s)\,.
\end{equation}
\end{enumerate}

\subsubsection{Point-wise values of wave}
Here we describe a "mechanism", which forms the values of waves
$u^f$.

Fix a $\gamma \in \Gamma$. In $\Omega \times [0,T]$ define the
{\it truncated} and {\it delayed hydras}
\begin{align}
\notag & H^T_\gamma:=\left\{(x,t)\in H_\gamma\,|\,\,0 \leqslant t
\leqslant T \right\}\,, \quad H^{T,s}_\gamma:= \\
\label{truc del hydras} &=\, \left\{(x,t+s)\in \Omega \times
[0,T]\,|\,\,(x,t)\in H_\gamma^T \right\},
\end{align}
where $s \in (0,T)$ is a {\it delay}. Also, we put
$H^{T,0}_\gamma:=H^T_\gamma$ and $H^{T,T}_\gamma:=(\gamma,T)$.
Each $H^{T,s}_\gamma$ is endowed with an amplitude function by
\begin{equation}\label{delayed amplitudes}
a^{T,s}(x,t)\,:=\,a(x, t-s)\,, \end{equation} where $a$ is the
amplitude on $H_\gamma$.

A set $$\kappa^{T,s}\,:=\, \left\{x \in \Omega\,|\,\,(x,T) \in
H^{T,s}_\gamma\right\}=\,{\rm supp\,}u^{\delta_\gamma
\delta}(\,\cdot\,, T-s)\qquad (0\leqslant s \leqslant T)$$
consists of finite number of points in $\Omega$, which we call
{\it heads} of the hydra $H^{T,s}_\gamma$.  The heads and delays
are related by
\begin{equation}\label{heads and delays}
\kappa^{T, s}\,=\,\pi\left(\rho^{-1}(T-s)\right)\,.
\end{equation}
The heads move into $\Omega$ as $s$ varies.

Fix a point $x \in \Omega$. We say that a hydra $H^{T,s}_\gamma$
influences on $x$ and write $$H^{T,s}_\gamma \rhd x$$ if $x \in
\kappa^{T,s}$, i.e., $x$ is one of the heads of $H^{T,s}_\gamma$.
The value $a^{T,s}(x,T)$ is referred to as amplitude of influence.
As is easy to see, for any $x \in \overline{\Omega^T[\gamma]}$
there is at least one hydra, which influences on $x$. The number
of hydras influencing on $x$ is always finite and equal to $\sharp
\rho\left(\pi^{-1}(x)\right)$.
\begin{Convention}\label{Conv 3}
Here and in what follows, dealing with the truncated hydra
$H^T_\gamma$, we understand $\pi^{-1}(A)$ as $\pi^{-1}(A)\cap
H^T_\gamma$.
\end{Convention}

Take a control $f \in {\cal F}^T_\gamma$ of the form
$f(\gamma',t)={\delta_\gamma}(\gamma')\varphi(t)$ with $\varphi
\in C[0,T]$. Fix an $x\in \Omega \backslash \Gamma$. A structure
of the fundamental solution and representation (\ref{Duhamel})
easily imply that the value $u^f(x,T)$ can be calculated by the
following procedure:
\begin{itemize}
\item find all $s \in [0,T)$ such that $H^{T,s}_\gamma \rhd x$ and
determine the corresponding amplitudes $a^{T,s}(x,T)$ \item find
\begin{equation}\label{value of u^f 1}
u^f(x,T)=\sum \limits_{s:\,\,H^{T,s}_\gamma \rhd
x}a^{T,s}(x,T)\,\varphi(s)=\sum \limits_{\sigma \in
\rho\left(\pi^{-1}(x)\right)}a(x,\sigma)\,\varphi(T-\sigma).
\end{equation}
\end{itemize}
For any $f \in {\cal F}^T$ of the form $f=\sum_{\gamma \in
\Gamma}\delta_\gamma(\,\cdot\,)\varphi_\gamma$ with
$\varphi_\gamma \in C[0,T]$, relation (\ref{value of u^f 1})
evidently implies
\begin{equation}\label{value of u^f 2}
u^f(x,T)=\sum \limits_{\gamma \in \Gamma}\sum \limits_{s:\,
H^{T,s}_\gamma \rhd x}a^{T,s}_\gamma(x,T)\,\varphi_\gamma(s)\,,
\end{equation}
where $a^{T,s}_\gamma(x,T)$ are the amplitudes on hydras $H^{T,
s}_\gamma$. Also, representing
$u^f(\,\cdot\,,t)=u^{f_{T-t}}(\,\cdot\,,T)$ by (\ref{steady
state}), one can find the value of the wave for any intermediate
$t \in (0,T)$ via (\ref{value of u^f 1}), (\ref{value of u^f 2}).

\subsection{Reachable sets}
\subsubsection{Definition}
In dynamical system (\ref{Graph 1})--(\ref{Graph 4}), a set of
waves
$${\cal U}^s\,:=\,
\left\{u^f(\,\cdot\,,s)\,|\,\, f \in {\cal F}^T\right\} \qquad (0
<s \leqslant T)$$ is said to be {\it reachable} (from the boundary
at the moment $t=s$). On graphs, ${\cal U}^s$ is a closed subspace
in $\cal H$. Its structure is of principal importance for many
applications, in particular to inverse problems: see
\cite{ContrGraph} - \cite{BV1}, \cite{LLS}.

By (\ref{F=sum F gamma}), we have
$$ {\cal U}^s\,=\,\sum \limits_{\gamma \in \Gamma}{\cal U}^s_\gamma$$
(algebraic sum), where
$${\cal U}^s_\gamma\,:=\,
\left\{u^f(\,\cdot\,,s)\,|\,\, f \in {\cal F}^T_\gamma\right\}
\qquad (0 <s \leqslant T)$$ are the sets reachable from single
boundary vertices.
\smallskip

By (\ref{steady state}), to study ${\cal U}^s_\gamma$ is to study
${\cal U}^T_\gamma$, and we deal mainly with the latter set. Its
structure will be described in detail. However, the description
requires certain preliminary considerations in 2.4.2 -- 2.4.4.

\subsubsection{Lattices and determination set}
We say two different points $l'=(x',t'),\, l''=(x'',t'')$ of
$H^T_\gamma$ to be {\it neighbors} and write $l'\simeq l''$, if
either $x'=x''$ or $t'=t''$ (equivalently: either
$\pi(l')=\pi(l'')$ or $\rho(l')=\rho(l'')$).

We write $l'\cong l''$ if there are $l_k$ such that $l'\simeq l_1
\simeq l_2\simeq \dots \simeq l_p \simeq l''$. As is easy to
check, $\cong$ is an equivalence on the hydra. For an $l \in
H^T_\gamma$, its equivalence class is called a {\it lattice} and
denoted by ${\cal L}[l]$ \footnote{Here we regard ${\cal L}[l]$ as
a subset of $H^T_\gamma$ but not as an element of the factor-set
$H^T_\gamma/\cong$.} .

For a $B \subset H^T_\gamma$ we set $${\cal L}[B]:=\bigcup_{l \in
B}{\cal L}[l] \subset H^T_\gamma.$$

We omit simple proofs of the following facts, which can be derived
from the above-accepted definitions:
\begin{itemize}
\item for any $B \subset H^T_\gamma$, one has
$\pi^{-1}\left(\pi\left({\cal L}[B]\right)\right) =
\rho^{-1}\left(\rho\left({\cal L}[B]\right)\right) ={\cal L}[B]$
\item the operation ${\cal L}: B \mapsto {\cal L}[B]$ satisfies
the Kuratovski axioms:

\noindent (i)\,\,\,({\it extensiveness})\,\,\,\, ${\cal L}[B]
\supset B$,

\noindent (ii)\,\,\,({\it idempotency})\,\,\,\, ${\cal L}[{\cal
L}[B]]={\cal L}[B]$

\noindent (iii)\,\,\,({\it additivity})\,\,\,\, ${\cal L}[B'\cup
B'']={\cal L}[B']\cup {\cal L}[B'']$

\noindent and, hence, is a topological closure. More precisely,
there is a unique topology on the hydra, in which the closure
coincides with $\cal L$ (see, e.g., \cite{Kelley}). \item each
${\cal L}[l]$ is a finite set; it is a closure of the single point
set $\{l\}$ in the above mentioned topology. Any point $l^\prime
\in {\cal L}[l]$ determines the whole set ${\cal L}[l]$.
\end{itemize}

For a point $x \in \overline{\Omega^T[\gamma]} \backslash \Gamma$,
define its {\it determination set} by
\begin{equation}\label{Lambda=pi(L)}
\Lambda_\gamma^T[x]\,:=\,\pi\left({\cal
L}[\pi^{-1}(x)]\right)\,\subset \Omega\,.
\end{equation}
The {\it alternating property} holds: for $x\not= x'$, one has
either $\Lambda_\gamma^T[x]=\Lambda_\gamma^T[x']$ or
$\Lambda_\gamma^T[x]\cap\Lambda_\gamma^T[x']=\emptyset$.

Determination set $\Lambda_\gamma^T[x]$ consists of the heads of
the delayed hydras $H^{T,s_i}_\gamma$, which satisfy $T-s_i \in
\rho\left({\cal L}\left[\pi^{-1}(x)\right]\right)$. It is the
hydras, which enter in representation (\ref{value of u^f 1}).
\smallskip
\begin{figure}
 \begin{center}
 \epsfysize=8cm
 \epsfbox{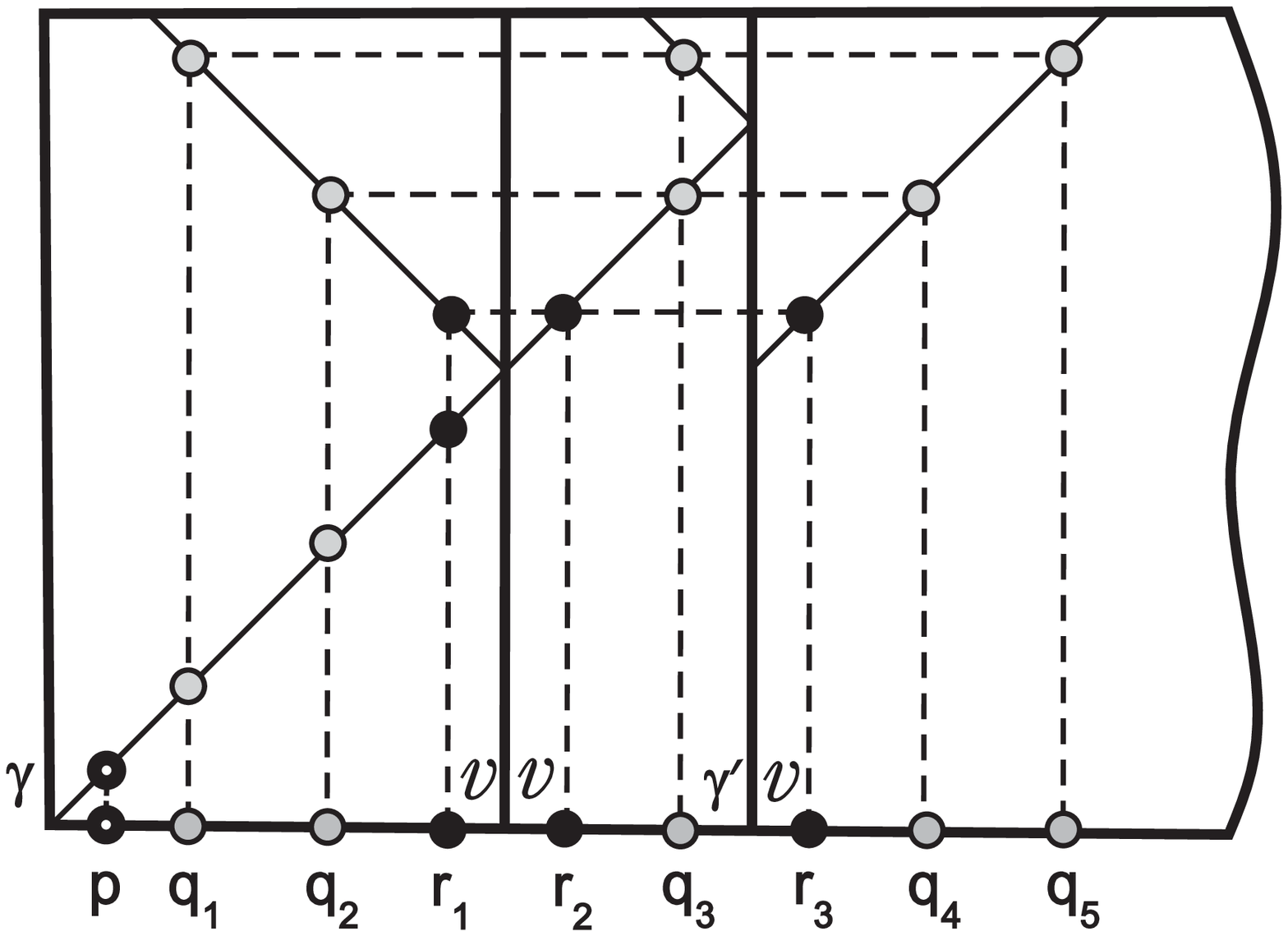}
 \end{center}
\caption{Lattices and sets $\Lambda_\gamma^T[x]$}
 \end{figure}

On Fig.5,
\begin{itemize}
\item ${\cal L}\left[\pi^{-1}(p)\right]$ is the point of
$H^T_\gamma$ above $p$, $\Lambda^T_\gamma[p]=\{p\}$

\item ${\cal L}\left[\pi^{-1}(q_m)\right]$ is the grey points on
$H^T_\gamma$, $\Lambda^T_\gamma[q_m]=\{q_1, q_2, q_3, q_4, q_5\}$

\item ${\cal L}\left[\pi^{-1}(r_m)\right]$ is the black points on
$H^T_\gamma$, $\Lambda^T_\gamma[r_m]=\{r_1, r_2, r_3\}$.
\end{itemize}

\subsubsection{Amplitude vectors}
Return to representation (\ref{value of u^f 1}) and modify it as
follows.

Let $$\rho\left({\cal
L}\left[\pi^{-1}(x)\right]\right)=\{t_i\}_{i=1}^N,\,\,\,0\leqslant
t_1<t_2<\dots<t_N\leqslant T\,,$$ so that $N=\sharp\rho\left({\cal
L}\left[\pi^{-1}(x)\right]\right)$. Introduce the functions
$\alpha^{T,t_i}: \Lambda_\gamma^T[x] \to {\mathbb R}$,
\begin{equation}\label{amplitude vectors}
\alpha^{T,t_i}(h)\,:=\,
 \begin{cases} a(h,t_i) & {\rm if}\,\,\,
(h,t_i) \in H^T_\gamma\\ 0 & {\rm otherwise}
 \end{cases}
\end{equation}
and call them {\it amplitude vectors}.
\smallskip

So, the set $\Lambda_\gamma^T[x]$ is endowed with amplitude
vectors $\alpha^{T,t_1}, \dots, \alpha^{T,t_N}$. Let ${\bf
l}_2(\Lambda_\gamma^T[x])$ be Euclidean space of functions on
$\Lambda_\gamma^T[x]$ with the standard product
$$\langle\alpha, \beta\rangle\,=\,\sum \limits_{h \in \Lambda^T[x]}
\alpha(h)\,\beta(h)\,.$$ By ${\cal A}^T[x]$ we denote the subspace
in ${\bf l}_2(\Lambda_\gamma^T[x])$ generated by amplitude
vectors:
$${\cal A}^T[x]\,:=\,{\rm span\,}\left\{\alpha^{T,t_1}, \dots,
\alpha^{T,t_N}\right\}\,.$$
\smallskip

Now, we are able to clarify the meaning of definition
(\ref{Lambda=pi(L)}) and the term "determination set". In
accordance with (\ref{value of u^f 1}), given $f=\delta_\gamma
\varphi$, for all $x' \in \Lambda_\gamma^T[x]$ the values
$u^f(x',T)$ {\it are determined} by hydras $H^{T,s_i}_\gamma$,
whose heads constitute $\Lambda_\gamma^T[x]$. Moreover, each value
is a combination of components of the amplitude vectors
$\alpha^{T,t_1}, \dots, \alpha^{T,t_N}$. Hence, one can regard the
set of values $\left\{u^f(x',T)\,|\,\, x' \in
\Lambda_\gamma^T[x]\right\}$ as an element of ${\bf
l}_2(\Lambda_\gamma^T[x])$, whereas (\ref{value of u^f 1}) can be
written in the form
\begin{equation}\label{u^f on Lambda}
u^f(\,\cdot\,,T)\big|_{\Lambda_\gamma^T[x]} = \sum \limits_{i=1}^N
\alpha^{T,t_i} \varphi(T-t_i)\,\in\, {\cal A}^T[x] .
\end{equation}
Also, one has to keep in mind the dependence
$t_i=t_i(x),\,\,N=N(x)$.

\subsubsection{Partition $\Pi^T_\gamma$}
Recall that $\Lambda_\gamma^T[x]$ is defined for points $x \in
\overline{\Omega^T[\gamma]} \backslash \Gamma$. For a while, let
$x$ belong to the open set ${\Omega^T[\gamma]} \backslash [\Gamma
\cup V]$; hence, there is an edge $e \ni x$.

Under the conditions, which we are going to specify now, small
variations of position of $x$ on $e$ lead to small variations of
the set $\Lambda_\gamma^T[x]$ in $\Omega$, which do not change its
"staff". Namely, the number $$M:=\sharp\Lambda_\gamma^T[x]= {\rm
dim\,}{\bf l}_2(\Lambda_\gamma^T[x])\,,$$ the number $N$ of
amplitude vectors $\alpha^{T,t_i}$, and values of the vectors
remain the same (do not depend on $x$). The question arises: What
are the bounds for such variations? In this section the answer is
given.
\smallskip

\noindent{\bf Critical points.}\,\,\,Recall that the corner points
on the complete hydra $H_\gamma$ are introduced at the end of
2.2.5. Dealing with $H^T_\gamma$, it is also convenient to assign
its top $\{(x,t) \in H^T_\gamma\,|\,\,t=T\}=\rho^{-1}(T)$ to
corner points. So, we say a space-time point $(x,t) \in
H^T_\gamma$ to be a {\it corner point} if it is a corner point of
$H_\gamma$ or belongs to the top of $H^T_\gamma$. The set of
corner points is denoted by ${\rm Corn\,}H^T_\gamma$. Note that,
for any vertex $w \in [V\cup \Gamma]\cap
\overline{\Omega^T[\gamma]}$, its pre-image $\pi^{-1}(w)$ consists
of corner points.

Introduce the lattice ${\cal L}[{\rm Corn\,} H^T_\gamma]\,,$ which
is a finite set of points on $H^T_\gamma$. This lattice divides
hydra $H^T_\gamma$ so that the set $H^T_\gamma \backslash {\cal
L}[{\rm Corn\,} H^T_\gamma]$ consists of a finite number of open
space-time intervals, which do not contain corner points. By the
latter, on these intervals the amplitude $a(\cdot)$ takes {\it
constant values}.

Points of the set \begin{equation}\label{critical points
gamma}\Theta^T_\gamma\,:=\,\pi\left({\cal L}\left[{\rm Corn\,}
H^T_\gamma\right]\right)\,\subset \overline{\Omega^T[\gamma]}
\end{equation}
are called {\it critical}.

Critical points divide neighborhood $\overline{\Omega^T[\gamma]}$
into parts. Namely, the set $\overline{\Omega^T[\gamma]}\backslash
\Theta^T_\gamma\,$
 is a collection of open intervals, each interval
lying into an edge of $\Omega$, whereas the critical points are
the endpoints of these intervals. We refer to this collection as a
{\it partition} $\Pi^T_\gamma$.
\smallskip

\noindent{\bf Families and cells.}\,\,\,Intervals of partition
$\Pi^T_\gamma$ are joined in the {\it families} as follows. Let
$c, c^\prime \in e$ be critical points such that the interval
$\omega =\,\,]c, c^\prime[\,\, \subset e$ contains no critical
points. This means that $\omega \in \Pi^T_\gamma$. As one can
easily see, the preimage $\pi^{-1}\left(\omega\right)$ consists of
a finite number of connected components. Each component is a
(space-time) interval on $H^T_\gamma$ of the same (space-time)
length $\sqrt{2}\, \tau(c, c')$ (see (\ref{metric on Char Set})),
the interval being free of corner points. By the latter, the same
is valid for the lattice ${\cal
L}\left[\pi^{-1}\left(\omega\right)\right]$: it is also a finite
collection of open intervals on $H^T_\gamma$ of length $\sqrt{2}\,
\tau(c, c')$, which do not contain corner points.

As a consequence, the set
\begin{equation}\label{Family Phi}\Phi\,:=\pi \left({\cal
L}\left[\pi^{-1}\left(\omega\right)\right] \right) \supset
\omega
\end{equation}
turns out to be a finite collection of open intervals $\omega_1,
\omega_2, \dots , \omega_M \, \subset
\overline{\Omega^T[\gamma]}\backslash \Theta^T_\gamma$ (with
$\omega$ among them) of the same length:
\begin{equation}\label{Phi=sum of cells}\Phi=\bigcup_{l=1}^M \,\omega_m\,, \quad {\rm
diam\,}\omega_m=\tau(c, c')=:\delta_\Phi\,. \end{equation}

\noindent We say this collection to be a {\it family}, intervals
$\omega_m$ are called {\it cells} of $\Phi$.
\smallskip

Comparing definitions (\ref{Lambda=pi(L)}) and (\ref{Family Phi}),
(\ref{Phi=sum of cells}), one can easily conclude the following.
For any $x \in \Phi$, the set $\Lambda_\gamma^T[x] \subset \Phi$
consists of the points $x_1, \dots, x_M$, each cell $\omega_m $
containing one (and only one) point $x_m$. Hence, we have
\begin{equation}\label{number M}
\sharp \Lambda_\gamma^T[x]\,=\,{\rm dim\,}{\bf
l}_2(\Lambda_\gamma^T[x])\,=\,M\,, \quad \Phi=\bigcup_{x \in
\omega}\Lambda_\gamma^T[x]
\end{equation}
as $x$ varies in any cell $\omega \subset \Phi$.
\smallskip

Starting with another interval $\omega' \not \subset \Phi$ bounded
by critical points, we get another family $\Phi'$, which has no
mutual cells or points with $\Phi$. Going on this way, we get a
finite set of families $\Phi^1, \Phi^2, \dots , \Phi^J$ and
conclude that partition $\Pi^T_\gamma$ corresponds to the
representation
\begin{equation}\label{Partition Pi}
\overline{\Omega^T[\gamma]}\backslash \Theta^T_\gamma\,=\,\bigcup
\limits_{j=1}^J \Phi^j=\bigcup \limits_{j=1}^J \bigcup
\limits_{m=1}^{M_j} \omega^{(j)}_m
\end{equation}
in the form of disjoint sums, $${\rm
diam\,}\omega^{(j)}_1=\dots={\rm
diam\,}\omega^{(j)}_{M_j}=\delta_{\Phi^j}$$ being valid.

\begin{figure}
 \begin{center}
 \epsfysize=10cm
 \epsfbox{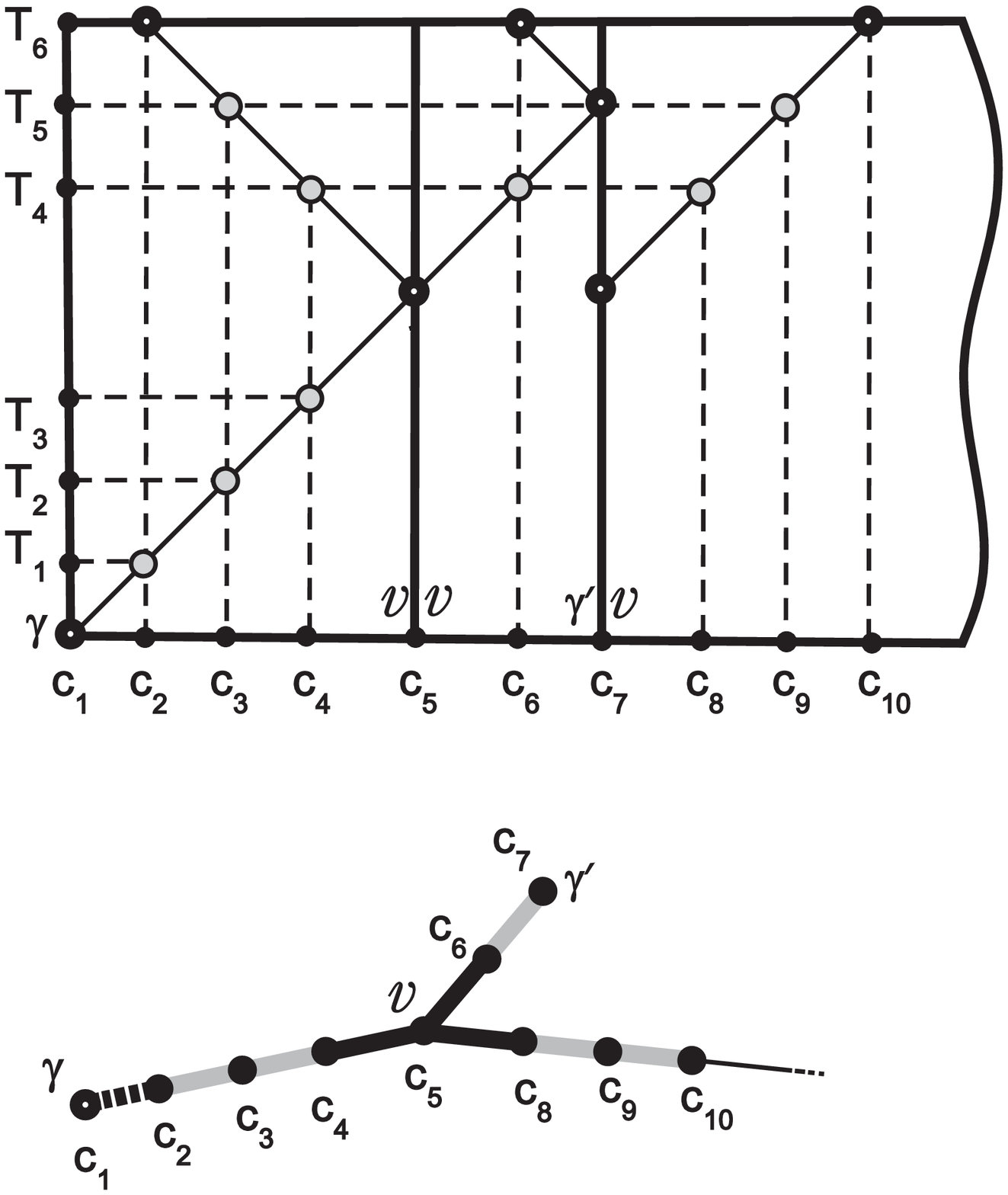}
 \end{center}
\caption{Partition $\Pi^T_\gamma[x]$}
 \end{figure}

On Fig.6,
\begin{itemize}
\item the set ${\rm Corn\,}H^T_\gamma$ is the black points with
holes, the lattice ${\cal L}\left[{\rm Corn\,}H^T_\gamma\right]$
is the grey points along with the corner points

\item the critical points set is
$\Theta^T_\gamma=\bigcup_{k=1}^{10}c_k,\,\,c_1=\gamma, c_5=v,
c_7=\gamma'$

\item the families and cells are
\begin{align*}
& \Phi^1=\omega_1^{(1)}=\,\,]c_1,c_2[ \quad(\rm dotted\,\, line),\\
&\Phi^2=\bigcup_{m=1}^5\omega_m^{(2)}=\,\,]c_2,c_3[\,\cup\,]c_3,c_4[\,\cup\,]c_6,c_7[\,\cup\,]c_8,
c_9[\,\cup\,]c_9,c_{10}[
\quad ({\rm grey\,\,intervals}),\\
&
\Phi^3=\bigcup_{m=1}^3\omega_m^{(3)}=\,\,]c_4,c_5[\,\cup\,]c_5,c_6[\,\cup\,]c_5,c_8[
\quad ({\rm black\,\,intervals})\,.
\end{align*}
\end{itemize}
\medskip

\noindent{\bf Variations and bounds.}\,\,\,Return to the question
on the bounds at the beginning of 2.4.5.

Take a cell $\omega =\,\,]c, c'[\,\,\subset \Phi=\bigcup
\limits^M_{l=1}\omega_m$ and choose an $x \in \omega$. The
determination set $\Lambda_\gamma^T[x]$ consists of the points
$x_1, \dots, x_M$ ($x$ among them), $x_m \in \omega_m$. Varying
$x$, one varies the set $\Lambda_\gamma^T[x]$ (the points $x_m$).

Parametrize
\begin{equation}\label{Parametrization x(r)}
\omega \ni x=x(r)\,, \qquad r=\tau(x,c) \in (0, \delta_\Phi);
\end{equation}
simultaneously, all $x_m(r) \in \Lambda_\gamma^T[x(r)]$ turn out
to be also parametrized. As $r$ varies from $0$ to $\delta_\Phi$,
each $x_m(r)$ runs over $\omega_m=\,]c_m, c_m'[$ (from $c_m$ to
$c_m'$ or in the opposite direction) and sweeps the cell
$\omega_m$. Correspondingly, $\Lambda_\gamma^T[x(r)]$ varies
continuously on the graph and sweeps the given family $\Phi$.

An important fact is that, in process of such varying, the
amplitude vectors $\alpha^{T,t_1\left(x(r)\right)}, \dots,
\alpha^{T,t_N\left(x(r)\right)} \in {\bf
l}_2(\Lambda_\gamma^T[x(r)])$ {\it do not vary}. This follows from
definition (\ref{amplitude vectors}): the points $\left(x_m(r),
t_i(x(r))\right)$ of the "horizontal layer" $\rho^{-1}(t_i)$ move
along the hydra but do not leave the intervals of $H^T_\gamma$, on
which there are no corner points, and hence amplitude $a$ takes
constant values (does not depend on $r$):
\begin{equation}\label{amplitude functions alpha i m}\alpha^{T,\,
t_i(x(r))}(x_m(r))\,=\,a\left(x_m(r),
t_i\left(x(r)\right)\right)\,=\, {\rm const}\,=:\,\alpha^{T,i}_m
\end{equation}
as $i=1, \dots, N;\,\,\,m=1, \dots, M$. Therefore, it is natural
to associate amplitude vectors not with the set
$\Lambda_\gamma^T[x]$ but the given family $\Phi \ni x$ and regard
them as piece-wise constant functions on $\Phi$ defined by
$$\alpha^{T,i}(x)\,:=\, \alpha^{T,i}_m \qquad {\rm as}\,\,\,x \in
\omega_m \subset \Phi\,.$$ We do it in what follows.
\medskip

If $x$ passes through a critical point $c$ and enters a cell of
another family, the picture of amplitude vectors changes. So, it
is the set $\Theta^T_\gamma$, which provides the bounds for
variations, which do not disturb ${\bf l}_2(\Lambda_\gamma^T[x])$
and ${\cal A}^T[x]$.
\smallskip

Varying $T$, one varies the set of critical points. Some of them
(in particular, the vertices $V\cup \Gamma$) do not change the
position in $\Omega$, the others are moving along the graph {\it
with velocity $1$}. By this, the set varies continuously in the
following sense: for a given $T$, there is a positive
$\varepsilon_0$ such that
\begin{equation}\label{Variation of crit points}
\Theta^{T+\varepsilon}_\gamma\,\subset \, \overline
{\Omega^{|\varepsilon|}[\Theta^{T}_\gamma]}
\end{equation}
holds for all $\varepsilon \in (-\varepsilon_0, \varepsilon_0)$.

\subsubsection{Local and global structure of wave}
Return to (\ref{u^f on Lambda}) and recall that, in such a general
representation, the amplitude vectors and delays are determined by
position of $x$:
\begin{equation*}
u^f(\,\cdot\,,T)\big|_{\Lambda_\gamma^T[x]}\, = \,\sum
\limits_{i=1}^{N(x)}\alpha^{T,t_i(x)}\,
\varphi\left(T-t_i(x)\right)\,.
\end{equation*}
Now, choose an $x$ in a cell $\omega$ of a family $\Phi=\bigcup
\limits^M_{m=1}\omega_m$ and parametrize by (\ref{Parametrization
x(r)}): $x_m(r) \in \omega_m,\,\,\,r \in (0,\delta_\Phi)$. Taking
into account (\ref{amplitude functions alpha i m}), we arrive at
basic representation of the wave $u^f$ {\it on the family} $\Phi$:
\begin{equation}\label{basic representation}
u^f\left(x_m(r), T\right)\, = \,\sum
\limits_{i=1}^{N}\alpha^{T,i}_m \,\psi_i(r)\,, \qquad r \in
(0,\delta_\Phi)\,,\,\,\,\,m=1, \dots, M\,,
\end{equation}
where $\psi_i(r):=\varphi\left(T-t_i\left(x(r)\right)\right)$.

By (\ref{Partition Pi}) and (\ref{basic representation}), we
represent the wave everywhere in $\Omega^T[\gamma]$ except of
critical points, i.e., almost everywhere on the graph. Such a
representation clarifies a local structure of waves in the cells
of families.

Recall that we deal with a control of the form
$f(\gamma',t)=\delta_\gamma(\gamma')\varphi(t)$ with $\varphi \in
C[0,T]$. However, one can extend the representation to all
controls $f \in {\cal F}^T_\gamma$ just by taking $\psi_i \in
L_2(0, \delta_\Phi)$ in (\ref{basic representation}).
\smallskip

Representation (\ref{basic representation}) provides a
characteristic description of waves on families. A function $y \in
{\cal H}=L_2(\Omega),\,\,\,{\rm supp\,}y \subset \overline \Phi$
is a wave (i.e., $y \in {\cal U}^T_\gamma$ does hold) if and only
if $y$ can be represented in the form of the right hand side of
(\ref{basic representation}) with $\psi_i \in L_2(0,
\delta_\Phi)$. Functions $\psi_i$ play the role of {\it
independent} function parameters, which determine a wave supported
in $\Phi$.

Taking into account (\ref{Partition Pi}), we get a global
characteristic description of the reachable set ${\cal
U}^T_\gamma$: to be a wave a function $y$ supported in
$\overline{\Omega^T[\gamma]}$ has to admit the representation
(\ref{basic representation}) on each $\Phi^j \subset
\Omega^T[\gamma]$.

\subsubsection{Projection $P^T_\gamma$} Let $P^T_\gamma$ be the
(orthogonal) projection in ${\cal H}=L_2(\Omega)$ onto the
reachable subspace ${\cal U}^T_\gamma$. Here we provide a
constructive description of this projection by the use of
representation (\ref{basic representation}).

For a subset $B \subset \Omega$, by $\chi_B$ we denote its {\it
indicator} (the characteristic function) and introduce the
subspace $${\cal H}\langle B \rangle\,:=\chi_B {\cal H}=\{\chi_B
y\,|\,\, y \in \cal H\}$$ of functions supported on $B$. In
accordance with (\ref{Partition Pi}) and results of 2.4.5, one has
\begin{equation}\label{H[Omega^T]= sum H <Phi>}
{\cal H}\langle \Omega^T[\gamma] \rangle=\oplus \sum \limits_{\Phi
\in \Pi^T_\gamma} {\cal H}\langle \Phi \rangle\,, \quad {\cal
U}^T_\gamma = \oplus \sum \limits_{\Phi \in \Pi^T_\gamma} {\cal
U}^T_\gamma\langle\Phi\rangle\,,
\end{equation}
where ${\cal U}^T_\gamma\langle\Phi\rangle \subset {\cal
H}\langle\Phi\rangle$ is the subspace of waves supported in $\Phi$
and represented by (\ref{basic representation}). Therefore,
\begin{equation}\label{P via Q Phi}
P^T_\gamma\,=\,\sum \limits_{\Phi \in \Pi^T_\gamma}Q_\Phi\,,
\end{equation}
where $Q_\Phi$ project in ${\cal H}\langle \Phi \rangle$ onto
subspaces ${\cal U}^T_\gamma\langle\Phi\rangle$. Hence, to
characterize $P^T_\gamma$ is to describe projections $Q_\Phi$.
\smallskip

Parametrize $\Phi$ by (\ref{Parametrization x(r)}) and introduce
an isometry $U$ by
\begin{equation}\label{def U Phi}
{\cal H}\langle \Phi \rangle \ni y \overset{U}\mapsto
\begin{pmatrix}
y(x_1(r))\\\dots\\y(x_M(r))
\end{pmatrix}\bigg|_{r \in (0, \delta_\Phi)} \in L_2\left((0, \delta_\Phi); {\mathbb
R}^M\right)\,.
\end{equation}
Since $\psi_i$ in (\ref{basic representation}) can be arbitrary,
this representation implies
$$U {\cal U}^T_\gamma\langle \Phi\rangle= L_2\left((0,
\delta_\Phi); {\mathbb A}_\Phi\right)\,, \quad
{\mathbb A}_\Phi:={\rm span\,}\{\begin{pmatrix} \alpha^{T, 1}_1\\\dots\\
\alpha^{T, 1}_M
\end{pmatrix}, \dots, \begin{pmatrix} \alpha^{T, N}_1\\\dots\\ \alpha^{T, N}_M
\end{pmatrix}\} \subset {\mathbb R}^M\,.$$
A substantial fact is that the vectors, which span ${\mathbb
A}_\Phi$, are constant (do not depend on $r$). Let
$p_\Phi=\{p_\Phi^{\,m m'}\}_{m,m'=1, \dots, M}$ be the (matrix)
projection in ${\mathbb R}^M$ onto ${\mathbb A}_\Phi$. Projection
$\tilde Q_\Phi$ in the space $L_2\left((0, \delta_\Phi); {\mathbb
R}^M\right)$ onto its subspace $L_2\left((0, \delta_\Phi);
{\mathbb A}_\Phi\right)$ acts {\it point wise} by the rule
$$(\tilde Q_\Phi v)(r)=\,p_\Phi \begin{pmatrix}
v_1(r)\\\dots\\v_M(r)
\end{pmatrix}, \qquad r \in (0, \delta_\Phi)\,.$$
In the mean time, one has $Q_\Phi= U^* \tilde Q_\Phi U$.
Summarizing, we arrive at the representation
\begin{equation}\label{representation}
\left(Q_\Phi y\right)\left(x_m(r)\right)\, = \,\sum
\limits_{m^\prime=1}^{M} p_\Phi^{\,m m^\prime}
y(x_{m^\prime}(r))\,, \quad r \in (0,\delta_\Phi)\,,\,\,\,\,m=1,
\dots, M
\end{equation}
with the {\it constant matrix} $p_\Phi$, which characterizes the
action of $Q_\Phi$.
\smallskip

\noindent{\bf System $\check \beta^T_\gamma$}\,\,\,The latter
representation can be written in more detail as follows.

Redesign the system of amplitude vectors $\{\alpha^{T,1}, \dots,
\alpha^{T,N}\}$ by the Schmidt procedure:
\begin{equation}\label{Schmidt}
\beta^{T,i}\,:=\,\begin{cases} \frac{\alpha^{T,i}-\sum
\limits_{j=1}^{i-1}\langle \alpha^{T,i},
\beta^{T,j}\rangle\,\beta^{T,j}}{\|\alpha^{T,i}-\sum
\limits_{j=1}^{i-1}\langle \alpha^{T,i},
\beta^{T,j}\rangle\,\beta^{T,j}\|}&{\rm if}\,\,\,\alpha^{T,i} \not
\in {\rm span\,}\{\alpha^{T,1}, \dots, \alpha^{T,i-1}\}\\ 0 & {\rm
otherwise}
\end{cases},
\end{equation}
and get a system ${\check \beta}^T:=\{\beta^{T,1}, \dots,
\beta^{T,N}\}$. Its nonzero elements satisfy $\langle \beta^{T,i},
\beta^{T,j} \rangle=\delta_{ij}$, and ${\rm span\,} \check \beta^T
= {\cal A}^T[x]$ holds.

By analogy with original vectors $\alpha^{T,i}$, it is convenient
to regard new amplitude vectors as piece-wise constant functions
on the family $\Phi$:
\begin{equation}\label{beta as functions}\beta^{T,i}(x)\,:=\,
\beta^{T,i}_m, \qquad x \in \omega_m \subset \Phi \,.
\end{equation}
Expressing the projection matrix $p_\Phi$ via system $\beta^{T,1},
\dots, \beta^{T,N}$ in (\ref{representation}), one can represent
the action of $Q_\Phi$ in the following final form
\begin{equation}\label{Q Phi via beta}
\left(Q_\Phi y\right)(x)\,=\,
\begin{cases}
\sum \limits_{i=1}^N \langle y\big|_{\Lambda_\gamma^T[x]},
\beta^{T,i}\rangle\,\beta^{T,i}(x)\,, & x \in \Phi\,,\\
0\,, & x \in \Omega \backslash \Phi
\end{cases}
\end{equation}
which is valid for any $y \in \cal H$.

At last, recalling (\ref{P via Q Phi}), we conclude that
$P^T_\gamma$ is characterized.

Note in addition that representations (\ref{P via Q Phi}), (\ref{Q
Phi via beta}) provide a look at controllability of a graph.
Recall a version of the {\it boundary control problem}: given $y
\in \cal H$ to find $f \in {\cal F}^T_\gamma$ such that
$u^f(\,\cdot\,,T)=y$ holds. {\it Controllability from $\gamma$}
means that ${\cal U}^T_\gamma={\cal H}$, i.e., this problem is
well solvable. In our terms, the latter is equivalent to the
relations ${\mathbb A}^T_{\Phi}={\mathbb R}^{M(\Phi)},\,\,\Phi \in
\Pi^T_\gamma$.
\medskip

\noindent{\bf Dependence on $T$.}\,\,\,Varying $T$, one varies the
neighborhood $\Omega^T[\gamma]$ filled with waves, reachable set
${\cal U}^T_\gamma$ and projection $P^T_\gamma$. As is evident,
${\cal U}^T_\gamma$ is increasing in $\cal H$ as $T$ grows. The
following arguments show that $P^T_\gamma$ is varied continuously.

Take a small $\Delta T>0$. The lattice ${\cal
L}\left[\rho^{-1}\left([T-\Delta T, T]\right)\right]\subset
H^T_\gamma$ is also "small": it consists of a final set of
(closed) intervals on the hydra, the total length of the intervals
vanishing as $\Delta T \to 0$. The same holds for the intervals in
$\Omega$, which constitute the set $\pi\left({\cal
L}\left[\rho^{-1}\left([T-\Delta T, T]\right)\right]\right)$. The
latter set is located in the small neighborhood $\Omega^{\Delta
T}[\Theta^T_\gamma]$ of critical points (see (\ref{Variation of
crit points})).

As is easy to see, for a point $x \not\in \pi\left({\cal
L}\left[\rho^{-1}\left([T-\Delta T, T]\right)\right]\right)$ one
has $\Lambda_\gamma^T[x]=\Lambda_\gamma^{T-\Delta T}[x]$, whereas
the amplitude vectors, which take part in projecting (\ref{basic
representation}), are the same: $\alpha^{T,i}=\alpha^{T-\Delta
T,i}$. Therefore, for any function $y \in \cal H$ we have
$\left(P^T_\gamma y\right)(x)=\left(P^{T-\Delta T}_\gamma
y\right)(x)$. Hence, the difference $P^T_\gamma y-P^{T-\Delta
T}_\gamma y$ has to be supported on the complement to such points:
\begin{equation}\label{supp (PT-P(T-d))y}
{\rm supp\,}\left(P^T_\gamma - P^{T-\Delta T}_\gamma \right) y
\,\subset \, \pi\left({\cal L}\left[\rho^{-1}\left([T-\Delta T,
T]\right)\right]\right) \subset \overline{\Omega^{\Delta
T}[\Theta^T_\gamma]}\,.
\end{equation}
As $\Delta T\to 0$, the neighborhood $\Omega^{\Delta
T}[\Theta^T_\gamma]$ shrinks to the finite set $\Theta^T_\gamma$
that implies $\|\left(P^T_\gamma - P^{T-\Delta T}_\gamma
\right)y\| \to 0$, i.e., $P^{T-\Delta T}_\gamma \to P^T_\gamma$ in
the strong operator topology in $\cal H$.

Quite analogous arguments with regard to property (\ref{Variation
of crit points}) show that $P^{T+\Delta T}_\gamma \to P^T_\gamma$
as $\Delta T \to 0$. Hence, $\{P^T_\gamma\}_{T \geqslant 0}$ is an
increasing {\it continuous} family of projections in $\cal H$.

\section{Eikonal algebra}
\subsection{Single eikonal}
Fix $\gamma \in \Gamma$ and $T>0$. Let $\Xi=\{\xi_k\}_{k=0}^K,
\,0=\xi_0<\xi_1< \dots <\xi_K=T$ be a partition of $[0,T]$ of the
range $r(\Xi)={\rm max\,}(\xi_k-\xi_{k-1})$; denote $\Delta
P^{\xi_k}_\gamma:=P^{\xi_k}_\gamma-P^{\xi_{k-1}}_\gamma$. With
each boundary vertex $\gamma \in \Gamma$ we associate a bounded
self-adjoint operator in $\cal H$ of the form
\begin{equation}\label{Eikonal def}
E^T_\gamma\,:=\, \int^T_0 \xi\,dP^\xi_\gamma\ = \lim
\limits_{r(\Xi)\to 0}\sum \limits_{k=1}^K \xi_k\,\Delta
P^{\xi_k}_\gamma
\end{equation}
(see, e.g., \cite{BSol}) and call it an {\it eikonal}. Our nearest
purpose is to describe how it acts.

\subsubsection{Small $T$}
Begin with the case $T \leqslant \tau (\gamma, V)$. In outer space
${\cal F}^T$, choose a control $f(\gamma',
t)=\delta_\gamma(\gamma') \varphi(t)$. By (\ref{first edge}) and
(\ref{Duhamel}), one has
$$u^f(x, t)= \varphi (t-\tau(x, \gamma))$$
(recall the Convention \ref{Conv 2}). Therefore, the reachable
sets are
$${\cal U}^\xi_\gamma=\{\varphi(\xi - \tau(\,\cdot\,,
\gamma))\,|\,\,\varphi \in L_2(0,T)\}= {\cal H}\langle
\Omega^\xi[\gamma]\rangle, \quad 0 \leqslant \xi \leqslant T\,.$$
Correspondingly, projection $P^\xi_\gamma$ in ${\cal H}$ onto
${\cal U}^\xi_\gamma$ cuts off functions on the part
$\Omega^\xi[\gamma]$ of the edge $e$ incident to $\gamma$, i.e.,
multiplies functions by the indicator $\chi_{\Omega^\xi[\gamma]}$.
Therefore, for a $y \in {\cal H}$, the summands in (\ref{Eikonal
def}) are
$$\left(\xi_k \Delta P^{\xi_k}_\gamma y\right)(x)\,=\,
\begin{cases}\xi_k\, y(x)\approx \tau(x, \gamma) y(x) & {\rm for}\,\,x \in \Omega^{\xi_k}[\gamma]
\backslash \Omega^{\xi_{k-1}}[\gamma]\\
0 & {\rm for \,\, other\,\,} x \in \Omega\end{cases}$$ ($\approx$
means that $\tau(x,\gamma)=\xi_k +O(r(\Xi))$ for $x \in
\Omega^{\xi_k}[\gamma] \backslash \Omega^{\xi_{k-1}}[\gamma]$).
Summing up the terms and passing to the limit as $r(\Xi)\to 0$, we
easily obtain
$$\left(E^T_\gamma y\right)(x)\,=\,
\begin{cases} \tau(x, \gamma)\, y(x) &  {\rm for}\,\,x \in
\Omega^T[\gamma]
\\
0 & {\rm for \,\, other\,\,} x \in \Omega\end{cases}\,.$$ Thus,
for small enough $T$'s, the eikonal cuts off functions on
$\Omega^T[\gamma]$ and multiplies by the distance to $\gamma$.

\subsubsection{Functions $\tau_i$}
Let $T>0$ be arbitrary. Choose a family $\Phi \in \Pi^T_\gamma$.
The set
\begin{equation}\label{sum of Ii}
\rho\left({\cal L}\left[\pi^{-1}(\Phi)\right]\right)\,=\,
\bigcup_{i=1}^N I_i
\end{equation}
consists of $N=N(\Phi)$ disjoined open intervals $I_i \subset
(0,T)$ of the same length $\delta_\Phi$. Note that the segments
$\overline{I_i}$ may intersect at the endpoints. On Fig.6, there
is $\rho\left({\cal L}\left[\pi^{-1}(\Phi^2)\right]\right)\,=\,
\bigcup_{i=1}^4I_i$, where $I_1=(T_1,T_2), I_2=(T_2, T_3),
I_3=(T_4,T_5), I_4=(T_5,T_6)$ (see also Fig.7).

With representation (\ref{sum of Ii}) one associates $N$ functions
on $\Phi$ of the form
\begin{align}
\notag & \tau_i(x)\,:=\,\begin{cases}
I_i\cap\rho\left(\pi^{-1}(x)\right) & {\rm
if}\,\,\,I_i\cap\rho\left(\pi^{-1}(x)\right)\not= \emptyset\\0 &
{\rm otherwise}
\end{cases}\,=\\
\label{Functions tau i} & =\begin{cases} t & {\rm
if}\,\,\,(x,t)\in \rho^{-1}\left(I_i\right) \\0 & {\rm otherwise}
\end{cases}\,,
\end{align}
which take values in the corresponding intervals $I_i$. They are
of clear geometric meaning in terms of the distance on the hydra:
if $\tau_i(x)\not=0$ then $(x, \tau_i(x)) \in H^T_\gamma$ and
\begin{equation}\label{distance along Hydra}
\tau_i(x)\,=\,\frac{1}{\sqrt{2}}\,\,\nu\left(\left(x,
\tau_i(x)),(\gamma,0\right)\right)
\end{equation}
holds (see (\ref{metric on Char Set})).
\smallskip

\begin{figure}
 \begin{center}
 \epsfysize=16cm
 \epsfbox{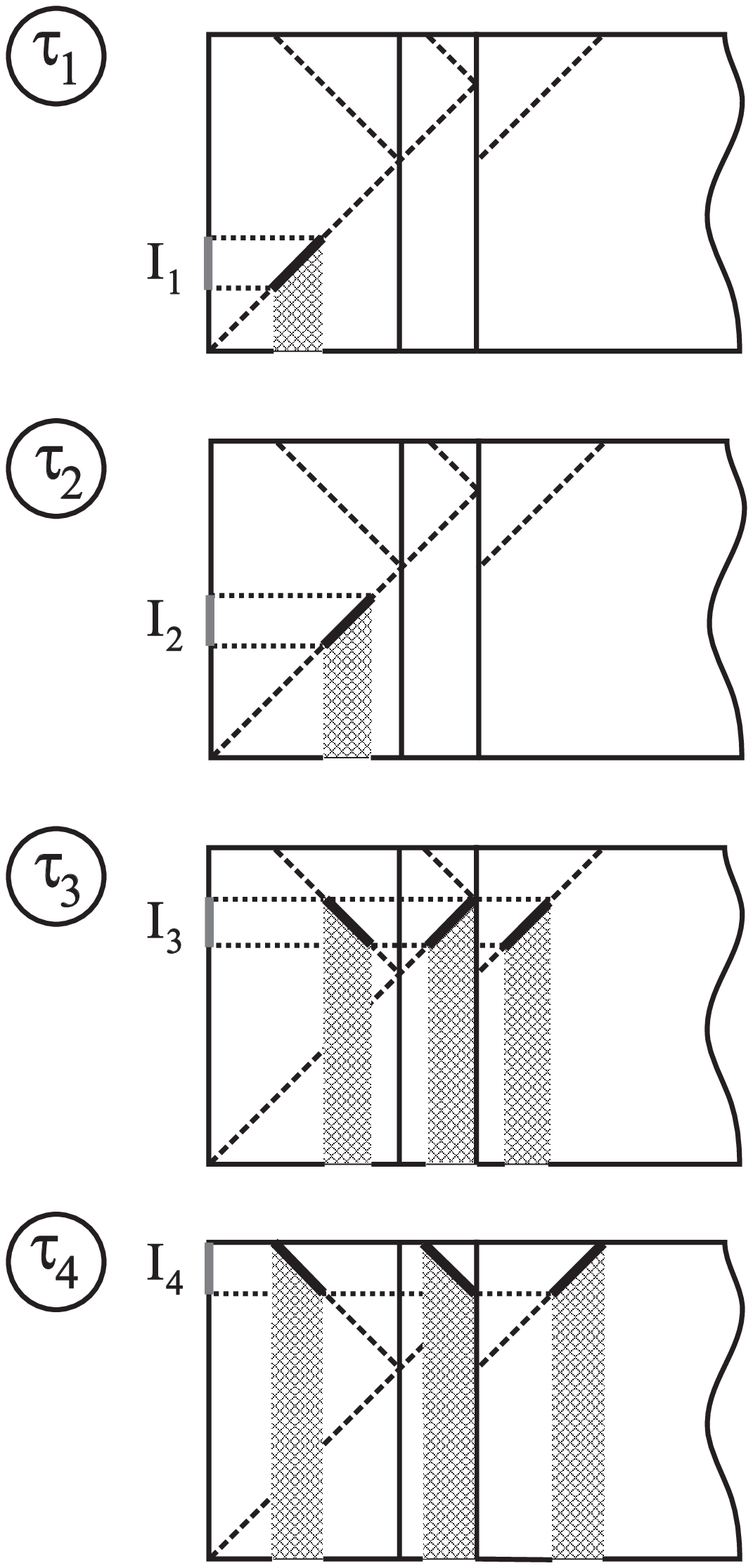}
 \end{center}
\caption{Functions $\tau_i$}
 \end{figure}

On Fig.7, the functions $\tau_1, \tau_2, \tau_3, \tau_4$, which
correspond to family $\Phi^2$ (see the grey part of $\Omega$ on
Fig.6), are shown. The space between the supports and graphs of
the functions is shaded.

\subsubsection{Arbitrary $T$}
Fix a $T>\tau(\gamma, V)$. Let $y \in {\cal H}$ be such that ${\rm
supp\,} y \subset \Phi \in \Pi^T_\gamma$ (i.e., $y \in {\cal
H}\langle \Phi \rangle$: see (\ref{H[Omega^T]= sum H <Phi>})).
Choose a $\xi \in (0,T]$ and a small $\Delta \xi>0$.
\smallskip

Assume that
$$ (\xi-\Delta \xi, \xi) \subset [0,T]\backslash \bigcup_{i=1}^N \overline
I_i\,.$$ In such a case one has $\pi\left({\cal
L}\left[\rho^{-1}\left((\xi-\Delta \xi,
\xi)\right)\right]\right)\cap \Phi=\emptyset$. Therefore, by
(\ref{supp (PT-P(T-d))y}), the supports of the functions
$(P^\xi_\gamma-P^{\xi-\Delta \xi}_\gamma)y$ and $y$ do not
intersect, and we have $(P^\xi_\gamma y - P^{\xi-\Delta
\xi}_\gamma y, y)=0$, i.e., $\|\Delta P^\xi_\gamma y\|=0$, where
$\Delta P^\xi_\gamma = P^\xi_\gamma -P^{\xi-\Delta \xi}_\gamma$.
By the latter, the interval $(\xi-\Delta \xi, \xi)$ contributes
nothing to the integral, which determines $E^T_\gamma y$.

By the aforesaid and with regard to continuity of $P^\xi_\gamma$,
for $y \in {\cal H}\langle \Phi \rangle$ the integral
(\ref{Eikonal def}) can be taken over the intervals $I_i$ only:
\begin{equation}\label{Eikonal 1}
E^T_\gamma y=\int_0^T \xi\,dP^\xi_\gamma y= \oplus
\sum_{i=1}^N\int_{I_i}\xi\,dP^\xi_\gamma y\,.
\end{equation}
The summands are pairwise orthogonal since the subspace
$P^\xi_\gamma \cal H$ is increasing as $\xi$ grows.
\smallskip

Let $\Phi$ and $y$ be the same as before. For what follows, it is
convenient to renumber the endpoints of the intervals so that
$I_i=(T_i, T_i+\delta_\Phi)$. The amplitude vectors $\beta^{T, i}$
are regarded as piece-wise constant functions on $\Phi$.

Now, assume that $(\xi-\Delta \xi, \xi) \subset I_1$. In this
case, the only amplitude vector, which contributes to the values
of $P^\xi_\gamma y$ on $\Phi$, is $\beta^{T, 1}$. Therefore, in
accordance with (\ref{Q Phi via beta}), (\ref{supp (PT-P(T-d))y})
we have
$$\left(\Delta P^\xi_\gamma y\right)(x)
\,=\, \Delta \chi^\xi(x)\langle y\big|_{\Lambda_\gamma^T[x]},
\beta^{T, 1} \rangle\, \beta^{T, 1}(x)\,, \qquad x \in \Phi\,,$$
where $\Delta \chi^\xi$ is the indicator of the set
$\pi\left({\cal L}\left[\rho^{-1}\left((\xi-\Delta \xi,
\xi)\right)\right]\right) \subset \Phi$. Correspondingly,
\begin{equation*}
\left(\xi \Delta P^\xi_\gamma y\right)(x) \,\approx\, \tau_1(x)
\Delta \chi^\xi(x)\langle y\big|_{\Lambda^T_\gamma[x]}, \beta^{T,
1} \rangle\,\beta^{T, 1}(x)\,, \qquad x \in \Phi\,,
\end{equation*}
where $\tau_i(x)$ are introduced by (\ref{Functions tau i}).
Summing up the terms of this form, one can easily justify the
limit passage as $r(\Xi)\to 0$ and get the equality
\begin{equation*}
\left(\int_{T_1}^{T_1+\delta_\Phi} \xi\, d P^\xi_\gamma
y\right)(x) = \tau_1(x) \langle y\big|_{\Lambda^T_\gamma[x]},
\beta^{T, 1} \rangle\,\beta^{T, 1}(x)\,, \qquad x \in \Phi\,.
\end{equation*}
\smallskip

Assume that $(\xi-\Delta \xi, \xi) \subset I_2$. In this case, the
amplitude vectors, which contribute to the values of $P^\xi_\gamma
y$ on $\Phi$, are $\beta^{T, 1}$ and $\beta^{T, 2}$. In the mean
time, (\ref{beta as functions}) and (\ref{Q Phi via beta}) imply
$P^{T_1+\delta_\Phi}_\gamma\beta^{T, 1}=\beta^{T, 1}$ that leads
to
\begin{align*}
& (\Delta P^\xi_\gamma y, \beta^{T, 1})_{{\cal H}\langle \Phi
\rangle}=(\Delta P^\xi_\gamma
y, P^{T_1+\delta_\Phi}_\gamma\beta^{T, 1})_{{\cal H}\langle \Phi \rangle}=\\
&\left(\left[P^\xi_\gamma P^{T_1+\delta_\Phi}_\gamma-
P^{\xi-\Delta \xi}_\gamma P^{T_1+\delta_\Phi}_\gamma\right]y,
\beta^{T, 1}\right)_{{\cal H}\langle
\Phi \rangle}= \\
& \left(\left[P^{T_1+\delta_\Phi}_\gamma-
P^{T_1+\delta_\Phi}_\gamma \right]y, \beta^{T, 1}\right)_{{\cal
H}\langle \Phi \rangle}\,=\,0
\end{align*}
by monotonicity of $P^\xi_\gamma$. Hence, $\Delta P^\xi_\gamma y$
has to be proportional to $\beta^{T,2}$ and we easily get
$$\left(\Delta P^\xi_\gamma y\right)(x)
\,=\, \Delta \chi^\xi(x)\langle y\big|_{\Lambda^T_\gamma[x]},
\beta^{T, 2}\rangle\,\beta^{T, 2}(x)\,, \qquad x \in \Phi\,.$$
Correspondingly,
\begin{equation*}
\left(\xi \Delta P^\xi_\gamma y\right)(x) \,\approx\, \tau_2(x)
\Delta \chi^\xi(x)\langle y\big|_{\Lambda^T_\gamma[x]},  \beta^{T,
2}, \rangle\,\beta^{T, 2}(x)\,, \qquad x \in \Phi\,.
\end{equation*}
Summing up such terms and passing to the limit, we obtain
\begin{equation*}
\left(\int_{T_2}^{T_2+\delta_\Phi} \xi\, d P^\xi_\gamma
y\right)(x) = \tau_2(x) \langle y\big|_{\Lambda^T_\gamma[x]},
\beta^{T, 2}\rangle\,\beta^{T, 2}(x)\,, \qquad x \in \Phi\,.
\end{equation*}
\smallskip

Continuing in the same way, with regard to (\ref{Eikonal 1}) we
arrive at the representation
\begin{equation}\label{Eikonal 2}
\left(E^T_\gamma y\right)(x)\,=\, \sum_{i=1}^N \tau_i(x) \langle
y\big|_{\Lambda^T_\gamma[x]}, \beta^{T, i}\rangle\,\beta^{T,
i}(x)\,, \qquad x \in \Phi\,.
\end{equation}
So, the eikonal projects functions $y \in {\cal
H}\langle\Phi\rangle$ on amplitude vectors and multiplies by
relevant distances.
\smallskip

Parametrize $\Phi$ by (\ref{Parametrization x(r)}), (\ref{def U
Phi}): $x_m(r) \in \omega_m \subset \Phi, \,\,\,m=1, 2, \dots, M,
\,\,\,\,0<r<\delta_\Phi$. Denote
\begin{align*} & \vec{y}(r):=\begin{pmatrix}
y(x_1(r))\\\dots\\y(x_M(r))
\end{pmatrix}, \quad B_\Phi:=\begin{pmatrix}
\beta^{T,1}_1 & \dots & \beta^{T,1}_M \\
\beta^{T,2}_1 & \dots & \beta^{T,2}_M \\
\dots & \dots & \dots \\
\beta^{T,N}_1 & \dots & \beta^{T,N}_M \\
\end{pmatrix}\,,\\
& D_\Phi(r)=\{\tau_i(r)\,\delta_{ij}\}_{i,j=1}^N\,,
\end{align*}
where $\tau_i(r):=\tau_i(x(r))$  is either $T_i+r$ or
$T_i+\delta_\Phi-r$ (see (\ref{Functions tau i})). Note that
$B_\Phi^*B_\Phi$ is the matrix of the projection $p_\Phi$ in
${\mathbb R}^M$ onto ${\mathbb A}_\Phi={\rm
span\,}\{{\beta}^{\,T,1}, \dots , {\beta}^{\,T,N}\}$. In this
notation, (\ref{Eikonal 2}) takes the form
\begin{equation}\label{Eikonal 3}
(\overrightarrow{E_\gamma y})(r)\,=\,[B_\Phi^* D_\Phi(r) B_\Phi]\,
\vec{y}(r)\,, \qquad r \in (0, \delta_\Phi)\,.
\end{equation}
\smallskip

Recalling the decomposition (\ref{H[Omega^T]= sum H <Phi>}), we
conclude that it reduces the eikonal and the representation
\begin{equation}\label{Eikonal }
E^T_\gamma\,=\,\oplus \sum \limits_{\Phi \in \Pi^T_\gamma}
E^T_\gamma \chi_\Phi
\end{equation}
holds, where $\chi_\Phi$ is understood as an operator in $\cal H$
multiplying by the indicator. Each part $E^T_\gamma \chi_\Phi$
acts in ${\cal H}\langle \Phi \rangle$ by (\ref{Eikonal 2}) (by
(\ref{Eikonal 3}) in the parametrized form).

Note in addition that (\ref{Eikonal 3}) and (\ref{Eikonal }) in
fact provide a canonical representation ({\it diagonalization}) of
the eikonal in the sense of the Spectral Theorem for self-adjoint
operators: see, e.g., \cite{BSol}. Also, one can easily see that
its spectrum $\sigma(E^T_\gamma)$ is ordinary (of multiplicity 1)
and absolutely continuous, $\sigma(E^T_\gamma)=[0,T]$.

\subsection{Algebra ${\mathfrak E}^T_\Sigma$}

\subsubsection{Definition}
Let $\Sigma \subseteq \Gamma$ be a subset of boundary vertices.
For controls $$f \in {\cal F}^T_\Sigma:=\oplus \sum_{\gamma \in
\Sigma}{\cal F}^T_\gamma\,,$$ the waves $u^f(\,\cdot\,,T)$ are
supported in the metric neighborhood $\overline{\Omega^T[\Sigma]}
\subset \Omega$ (see (\ref{supp u^f})). These waves constitute a
reachable set $${\cal U}^T_\Sigma\,=\,\sum_{\gamma \in
\Sigma}{\cal U}^T_\gamma\,\subset {\cal
H}\langle\Omega^T[\Sigma]\rangle$$ (algebraic sum).
\smallskip

Let ${\mathfrak B}(\cal H)$ be the normed algebra of bounded
operators in ${\cal H}$. With each $\gamma \in \Sigma$ one
associates the eikonal $E^T_\gamma \in {\mathfrak B}(\cal H)$. By
${\mathfrak E}^T_\Sigma$ we denote the {\it C*-subalgebra of
${\mathfrak B}(\cal H)$ generated by eikonals}
$\{E^T_\gamma\}_{\gamma \in \Sigma}$, i.e., the minimal
norm-closed C*-subalgebra in ${\mathfrak B}(\cal H)$, which
contains all these eikonals \cite{Dix}, \cite{Mur}.
\smallskip

Our paper is written for the sake of introducing algebra
${\mathfrak E}^T_\Sigma$. It is defined by perfect analogy with
the eikonal algebras associated with Riemannian manifolds: see
\cite{BSobolev}, \cite{BD_2}. In the rest of the paper, we clarify
a structure of ${\mathfrak E}^T_\Sigma$.

\subsubsection{Partition $\Pi^T_\Sigma$}
The set
$$H^T_\Sigma \,:=\,
\bigcup \limits_{\gamma \in \Sigma}H^T_\gamma \,\subset\,
{\Omega^T[\Sigma]} \times [0,T]$$ is also said to be a {\it
hydra}. It is also a space-time graph.
\smallskip

The analogs of the objects, which are related with $H^T_\gamma$,
are introduced for $H^T_\Sigma$ as follows.
\begin{itemize}
\item The projections $\pi: (x,t) \mapsto x$ and $\rho: (x,t)
\mapsto t$ are now understood as the maps from $H^T_\Sigma$ to
$\Omega$ and $[0,T]$ respectively.

\item An {\it amplitude} on $H^T_\Sigma$ is
$$a(x,t)\,:=\,\sum \limits_{\gamma \in \Sigma:\,\,\,(x,t)\in H^T_\gamma}\,a_\gamma(x,t)\,,$$
where $a_\gamma$ is the amplitude on $H^T_\gamma$.

\item The equivalence $l' \cong l''$ and lattices ${\cal L}[B]$
are defined as in 2.4.2, just replacing $H^T_\gamma$ by
$H^T_\Sigma$.

\item The set ${\rm Corn\,}H^T_\Sigma$ of {\it corner points} is
defined as in 2.4.4, replacing $H^T_\gamma$ by $H^T_\Sigma$. The
evident relation
\begin{equation*}\label{corner points H Sigma}
{\rm Corn\,}H^T_\Sigma\, \supset \,\bigcup_{\gamma \in \Sigma}{\rm
Corn\,}H^T_\gamma
\end{equation*}
holds. However, the latter sum can be smaller than ${\rm
Corn\,}H^T_\Sigma$ because additional corner points on
$H^T_\Sigma$ do appear owing to intersection of the space-time
edges of $H^T_{\gamma}$ with edges of $H^T_{\gamma'}$ for
different $\gamma, \gamma' \in \Sigma$. It happens as
$T>\frac{1}{2}\,\tau(\gamma, \gamma')$.

\item  {\it Critical points} are introduced in the same way as
(\ref{critical points gamma}): they constitute a finite set
\begin{equation*}
\Theta^T_\Sigma\,:=\,\pi\left({\cal L}\left[{\rm
Corn\,}H^T_\Sigma\right]\right)\,\supset \bigcup_{\gamma \in
\Sigma}\Theta^T_\gamma\,. \end{equation*} \item Fix an $x \in
\overline{\Omega^T[\Sigma]}\backslash \Theta^T_\Sigma$; let
$]c,c'[\,\ni x$ be an open interval in the graph between the
critical points $c, c'$, which contains no critical points. By
analogy with (\ref{Lambda=pi(L)}), define a {\it determination
set}
$$\Lambda^T_\Sigma[x]\,:=\,\pi\left({\cal L}\left[\pi^{-1}(x)\right] \right)\,\ni\,x$$
(here $\cal L$ is a lattice on $H^T_\Sigma$!). As is evident, one
has $\Lambda^T_\Sigma[x] \supset \Lambda^T_\gamma[x]$ as $\gamma
\in \Sigma$. If $x$ runs over $]c, c'[$ from $c$ to $c'$, the set
$\Lambda^T_\Sigma[x]$ sweeps a {\it family} $\Phi=\bigcup_{m=1}^M
\omega_m$ of open intervals $\omega_m$ ({\it cells}) of the same
length $\delta_\Phi$.
\end{itemize}

By the aforesaid and analogy with (\ref{Partition Pi}), we have
the representation
\begin{equation}\label{Partition Pi Sigma}
\overline{\Omega^T[\Sigma]}\backslash \Theta^T_\Sigma\,=\,\bigcup
\limits_{j=1}^J \Phi^j=\bigcup \limits_{j=1}^J \bigcup
\limits_{m=1}^{M_j} \omega^{(j)}_m
\end{equation}
in the form of disjoint sums, where ${\rm
diam\,}\omega^{(j)}_1=\dots={\rm
diam\,}\omega^{(j)}_{M_j}=\delta_{\Phi^j}$. This representation is
referred to as a {\it partition} $\Pi^T_\Sigma$.

\begin{figure}
 \begin{center}
 \epsfysize=11cm
\epsfxsize=8cm
 \epsfbox{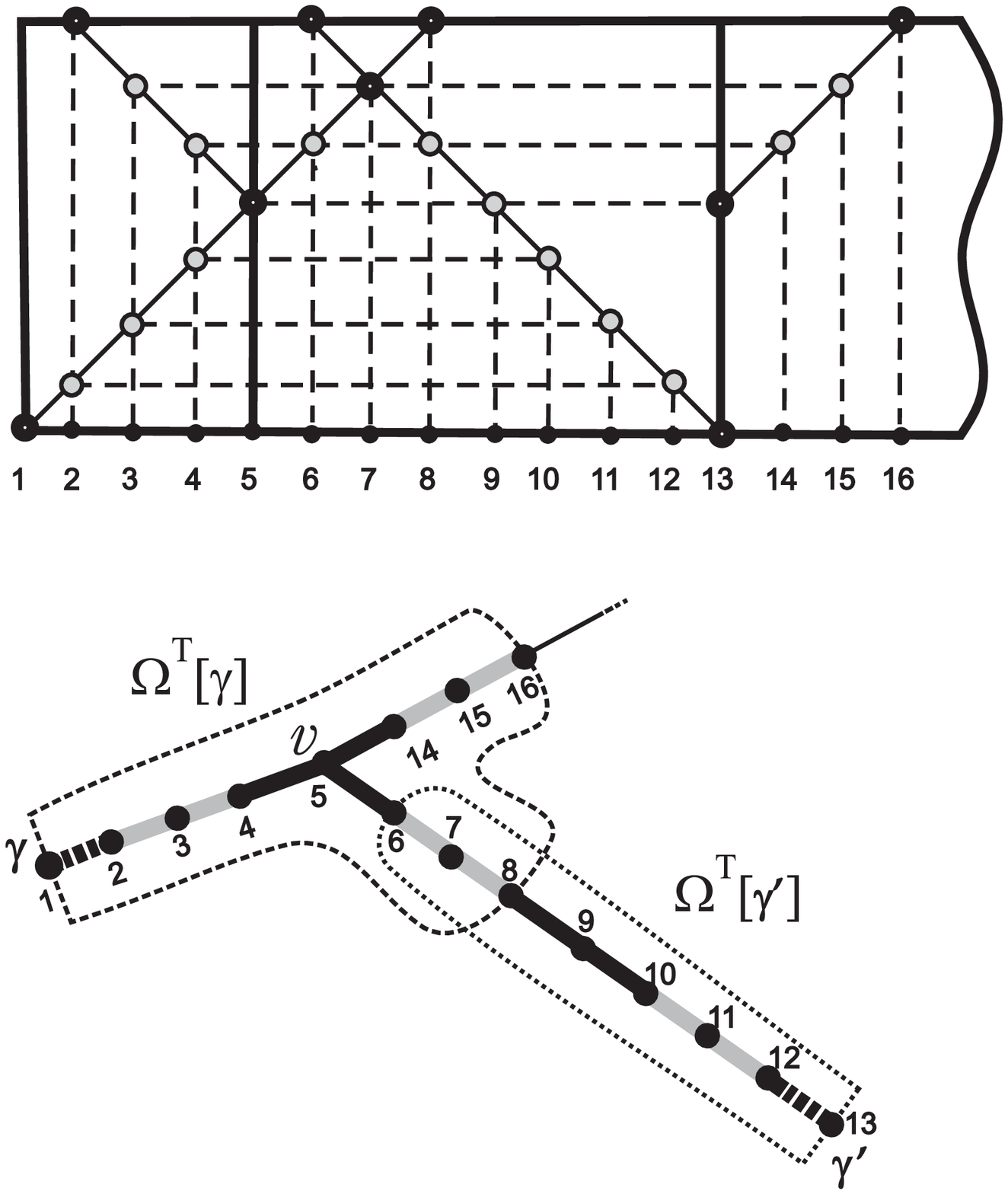}
 \end{center}
\caption{Partition $\Pi^T_\Sigma$}
 \end{figure}

Fig.8 illustrates the case $\Sigma=\{\gamma, \gamma'\}$ for
$T=\tau(\gamma, \gamma')+\varepsilon$:
\begin{itemize}
\item ${\rm Corn\,}H^T_\Sigma$ is the black points with small
holes at the center, ${\cal L}\left[{\rm Corn\,}H^T_\Sigma\right]$
is ${\rm Corn\,}H^T_\Sigma$ plus the grey points

\item the critical points
$\Theta^T_\Sigma=\bigcup_{k=1}^{16}c_k,\,\,c_1=\gamma, c_5=v,
c_{13}=\gamma'$ are denoted by $c_k\equiv k$

\item the families and cells are
\begin{align*}
\Phi^1=\bigcup_{m=1}^2\omega_m^{(1)} \,\,\,(\rm dashed),\quad
\Phi^2=\bigcup_{m=1}^8\omega_m^{(2)} \,\,\,({\rm grey}),
\,\,\,\Phi^3=\bigcup_{m=1}^5\omega_m^{(3)}\,\,\,({\rm black})
\end{align*}

\item the neighborhoods $\Omega^T[\gamma]$ and $\Omega^T[\gamma']$
filled with waves are contoured with the dashed lines.
\end{itemize}

\subsubsection{Functions $\tau_i^{\gamma, \Phi}$ and system
$\check \beta^T_{\gamma, \Phi}$} Choose a family
$\Phi=\bigcup_{m=1}^{M} \omega_m$, which is an element of the
partition $\Pi^T_\Sigma$. Quite analogously to (\ref{sum of Ii}),
the set
\begin{equation}\label{sum of Ii GENERAL}
\rho\left({\cal
L}\left[\pi^{-1}(\Phi)\right]\right)\,=\,\bigcup_{i=1}^N I_i
\end{equation}
consists of $N=N(\Phi)$ disjoined open intervals $I_i=(T_i,
T_i+\delta_\Phi) \subset (0,T)$ of the same length $\delta_\Phi$.
We number them so that $0 \leqslant T_1 < T_2 < \dots T_N <
T_N+\delta_\Phi \leqslant T$.
\smallskip

Fix a $\gamma \in \Sigma$. With representation (\ref{sum of Ii
GENERAL}) one associates $N$ functions on $\Phi$ of the form
\begin{align}
\notag & \tau^{\gamma, \Phi}_i(x)\,:=\,\begin{cases}
I_i\cap\rho\left(\pi^{-1}(x)\cap H^T_\gamma \right) & {\rm
if}\,\,\,I_i\cap\rho\left(\pi^{-1}(x)\cap H^T_\gamma \right)\not=
\emptyset\\0 & {\rm otherwise}
\end{cases}\,=\\
\label{Functions tau gamma i} & =\,\begin{cases} t & {\rm
if}\,\,\,(x,t)\in \rho^{-1}\left(I_i\right)\cap H^T_\gamma  \\0 &
{\rm otherwise}
\end{cases}\,.
\end{align}
As is easy to recognize, these functions are just a version of the
functions (\ref{Functions tau i}), the version being relevant to
partition $\Pi^T_\Sigma$.
\smallskip

For the family $\Phi \in \Pi^T_\Sigma$, a relevant version of the
system $\check \beta^T$ (see (\ref{Schmidt})) is constructed as
follows:
\begin{itemize}
\item For each $i=1, \dots, N$, take a $t_i \in I_i$ and choose an
$x \in \pi\left(\rho^{-1}(t_i)\right) \subset
\Lambda^T_\Sigma[x]=\{x_1, \dots, x_M\}$, where $x_m \in
\omega_m$. Define a vector $\alpha^i \in {\bf
l}_2(\Lambda^T_\Sigma[x])$ by
$$ \alpha^i(x_m)\,:=\,\begin{cases} a_\gamma(x_m, t_i) & {\rm as\,}\,\,(x_m,
t_i)\in H_\gamma^T\\ 0 & {\rm otherwise}
\end{cases}\,.$$
\item Redesign the system $\alpha^1, \dots, \alpha^N$ by the
Schmidt process (see (\ref{Schmidt})) and get the system $\check
\beta^T_{\gamma, \Phi}:=\{\beta^{T,1}_{\gamma, \Phi}, \dots,
\beta^{T,N}_{\gamma, \Phi}\}$. The amplitude subspace is
$${\cal A}^T_{\gamma, \Phi}[x]\, := \,{\rm span\,}\check
\beta^T_{\gamma, \Phi} \subset {\bf l}_2(\Lambda^T_\Sigma[x])\,.$$
\end{itemize}
Also, with each vector $\beta^{T,i}_{\gamma, \Phi}$ we associate a
piece-wise constant function \begin{equation}\label{beta as
functions Sigma} \beta^{T,i}_{\gamma, \Phi}(x)\,:=\,
(\beta^{T,i}_{\gamma, \Phi})_m, \qquad x \in \omega_m \subset \Phi
\end{equation}
(see (\ref{beta as functions})).

\subsubsection{Projections and eikonals}
Recall that $P^T_\gamma$ projects in $\cal H$ onto ${\cal
U}^T_\gamma$. Repeating the arguments, which have led to
representations (\ref{P via Q Phi}) and {(\ref{Q Phi via beta})},
one can modify them to the following form relevant to the complete
hydra $H^T_\Sigma$:
\begin{equation}\label{P via Q Phi Sigma}
P^T_\gamma \,= \,\oplus \sum \limits_{\Phi \in \Pi^T_\Sigma}
Q^\gamma_\Phi\,,
\end{equation}
where
\begin{equation}\label{Q Phi via beta Sigma}
\left(Q^\gamma_\Phi y\right)(x)\,=\,\begin{cases}\sum
\limits_{i=1}^{N(\Phi)} \left\langle
y\big|_{\Lambda_\Sigma^T[x]}\,,\, \beta^{T,i}_{\gamma,
\Phi}\right\rangle\,  \beta^{T,i}_{\gamma, \Phi}(x)\,,
  & x \in \Phi\\0, & x \in \Omega \backslash \Phi\end{cases} \,.
\end{equation}
\smallskip

Quite analogously, a relevant version of representations
(\ref{Eikonal 2}) and (\ref{Eikonal }) takes the form
\begin{equation}\label{Eikonal gamma Sigma}
E^T_\gamma\,=\,\oplus \sum \limits_{\Phi \in \Pi^T_\Sigma}
E^T_\gamma \chi_\Phi\,,
\end{equation}
where
\begin{equation}\label{Eikonal 2 GENERAL Sigma}
\left(E^\gamma_\Phi y\right)(x)\,=\,\begin{cases}\sum
\limits_{i=1}^{N(\Phi)} \tau^{\gamma, \Phi}_i(x)\,\left\langle
y\big|_{\Lambda_\Sigma^T[x]}\,,\, \beta^{T,i}_{\gamma,
\Phi}\right\rangle\, \beta^{T,i}_{\gamma, \Phi}(x)\,,
  & x \in \Phi\\0, & x \in \Omega \backslash \Phi\end{cases} \,.
\end{equation}
\smallskip

Fix a $\gamma \in \Sigma$ and choose a family
$\Phi=\bigcup^M_{m=1}\omega_m \in \Pi^T_\Sigma$; recall that the
number $N$ is defined in (\ref{sum of Ii GENERAL}). Parametrize
$\Phi$ by (\ref{Parametrization x(r)}), (\ref{def U Phi}): $x_m(r)
\in \omega_m, \,\,\,m=1, 2, \dots, M, \,\,\,\,0<r<\delta_\Phi$.
Denote
\begin{align*} & \vec{y}(r):=\begin{pmatrix}
y(x_1(r))\\\dots\\y(x_M(r))
\end{pmatrix}, \quad B_{\gamma, \Phi}:=\begin{pmatrix}
(\beta^{T,1}_{\gamma,\Phi})_1 & \dots & (\beta^{T,1}_{\gamma,\Phi})_M \\
(\beta^{T,2}_{\gamma,\Phi})_1 & \dots & (\beta^{T,2}_{\gamma,\Phi})_M \\
\dots & \dots & \dots \\
(\beta^{T,N}_{\gamma,\Phi})_1 & \dots &
(\beta^{T,N}_{\gamma,\Phi})_M
\end{pmatrix}\,,\\
& D_{\gamma,\Phi}(r)=\{D^{ij}_{\gamma,\Phi}(r)\}_{i,j=1}^N\,:
\quad D^{ij}_{\gamma,\Phi}(r)= \tau^{\gamma,\Phi}_i(r)\,
\delta_{ij}\,,
\end{align*}
where $\tau^{\gamma,\Phi}_i(r):=\tau^{\gamma,\Phi}_i(x(r))$  is
either $T_i+r$ or $T_i+\delta_\Phi-r$. Note that $B_{\gamma,
\Phi}^*B_{\gamma, \Phi}$ is the matrix of the projection
$p_{\gamma,\Phi}$ in ${\mathbb R}^M$ onto ${\mathbb
A}^T_{\gamma,\Phi}={\rm span\,}\check \beta^T_{\gamma, \Phi} $. In
this notation, the first line in the right hand side of
(\ref{Eikonal 2 GENERAL Sigma}) is
\begin{equation}\label{Eikonal 3 GENERAL Sigma}
(\overrightarrow{E^T_\gamma y})(r)\,=\,[B_{\gamma, \Phi}^*
D_{\gamma,\Phi}(r) B_{\gamma, \Phi}]\, \vec{y}(r)\,, \quad r \in
(0, \delta_\Phi)\,\,\,\,\,\,\,(\gamma \in \Sigma)\,,
\end{equation}
which is just a relevant form of (\ref{Eikonal 3}).
\smallskip

A key feature of representations (\ref{Eikonal 2 GENERAL Sigma})
and (\ref{Eikonal 3 GENERAL Sigma}) is the following. They
represent eikonals $E^T_\gamma$ in the form, which is common to
all the vertices $\gamma \in \Sigma$ and available for any family
$\Phi$ of partition $\Pi^T_\Sigma$ of the complete hydra
$H^T_\Sigma$.

\subsubsection{Block algebras. Structure of ${\mathfrak E}^T_\Sigma$.}
In accordance with (\ref{Partition Pi Sigma}), we have the
decomposition
\begin{equation}\label{BASIC decomposition}
{\cal H}\langle \Omega^T[\Sigma]\rangle\,=\,\oplus \sum
\limits_{\Phi \in \Pi^T_\Sigma}{\cal H}\langle\Phi\rangle\,,
\end{equation}
which reduces each eikonal $E^T_\gamma$ for $\gamma \in \Sigma$:
\begin{equation}\label{Eikonal GENERAL} E^T_\gamma\,=\,\oplus \sum
\limits_{\Phi \in \Pi^T_\Sigma} E^T_\gamma \big|_{{\cal
H}\langle\Phi\rangle}
\end{equation}
(compare with (\ref{H[Omega^T]= sum H <Phi>}) and (\ref{Eikonal
})). Each part $E^T_\gamma \big|_{{\cal H}\langle\Phi\rangle}$
acts in the subspace ${\cal H}\langle\Phi\rangle$ by (\ref{Eikonal
2 GENERAL Sigma}) or, equivalently, by (\ref{Eikonal 3 GENERAL
Sigma}) in the parametrized form.
\smallskip

Let ${\mathfrak b}^T_\Phi \subset {\mathfrak B}\left({\cal
H}\langle\Phi\rangle\right)$ be the C*-subalgebra generated by the
system\\ $\{E^T_\gamma\big|_{{\cal
H}\langle\Phi\rangle}\,|\,\,\gamma \in \Sigma\}$ of the eikonal
parts. We say ${\mathfrak b}^T_\Phi$ to be a {\it block algebra}.
\smallskip

By (\ref{Eikonal 3 GENERAL Sigma}), each ${\mathfrak b}^T_\Phi$ is
isometrically isomorphic to the subalgebra $\tilde{\mathfrak
b}^T_\Phi \subset \\{\mathfrak B}\left(L_2\left((0,\delta_\Phi);
{\mathbb R}^{M(\Phi)}\right)\right)$ generated by the operators,
which multiply elements (vector-functions) $\vec{y}(\cdot)$ by the
matrix-functions $B_{\gamma, \Phi}^* D_{\gamma,\Phi}(\cdot)
B_{\gamma, \Phi}$\,\,($\gamma \in \Sigma$). These functions are
continuous \footnote{moreover, the matrix elements are the linear
functions of $r \in [0,\delta_\Phi]$}, and hence we have
\begin{equation}\label{structure of E}\tilde{\mathfrak b}^T_\Phi\,\subset
C\left(\left[0,\delta_\Phi\right]; {\mathbb
M}^{M(\Phi)}\right)\,,\end{equation} where the latter is the
algebra of continuous real $M(\Phi) \times M(\Phi)$ - matrix
valued functions on $[0, \delta_\Phi]$.
\medskip

Just summarizing these considerations, we arrive at the main
result of the paper: decomposition (\ref{BASIC decomposition})
reduces eikonal algebra ${\mathfrak E}^T_\Sigma$, and the
representation
\begin{equation}\label{BASIC}
{\mathfrak E}^T_\Sigma\,=\,\oplus \sum_{\Phi \in
\Pi^T_\Sigma}{\mathfrak b}^T_\Phi
\end{equation}
holds.

\subsubsection{Noncommutativity}
As we noted in Introduction, algebra ${\mathfrak E}^T_\Sigma$ is
noncommutative. The reason is that the matrix projections
$p_{\gamma, \Phi}$ (in ${\mathbb R}^{M(\Phi)}$ onto ${\mathbb
A}^T_{\gamma,\Phi}$) corresponding to different $\gamma \in
\Sigma$ do not have to commute.  As a result, eikonal parts
$E^T_\gamma|_{{\cal H}\langle\Phi\rangle}$ and
$E^T_{\gamma^\prime}|_{{\cal H}\langle\Phi\rangle}$ do not commute
as $\gamma \not=\gamma^\prime$, so that the block-algebra
${\mathfrak b}^T_\Phi$ turns out to be noncommutative.

Moreover, in a sense, the eikonal algebra on a graph is {\it
strongly} noncommutative. We mean the following. In the Maxwell
dynamical system on a Riemannian manifold, a straightforward
analog of ${\mathfrak E}^T_\Sigma$ is also a noncommutative
algebra but its factor over the ideal of compact operators turns
out to be commutative \cite{BD_2} (and, moreover, isometric to the
algebra $C(\Omega)$). This may be referred to as a {\it weak}
noncommutativity. On graphs, it is definitely not the case: simple
examples show that no factorization eliminates noncommutativity of
a generic block-algebra.

\subsubsection{Spectrum}
A {\it spectrum} $\widehat{\mathfrak E}^T_\Sigma$ of the algebra
${\mathfrak E}^T_\Sigma$ is the set of its primitive ideals
endowed with the Jacobson topology (see \cite{Dix}, \cite{Mur}).
By (\ref{BASIC}), one has
\begin{equation}\label{spec E= sum spec b}
\widehat{\mathfrak E}^T_\Sigma\,=\,\bigcup_{\Phi \in
\Pi^T_\Sigma}\widehat{\mathfrak b}^T_\Phi\,,
\end{equation}
so that to study a structure of $\widehat{\mathfrak E}^T_\Sigma$
is to analyze $\widehat{\mathfrak b}^T_\Phi$. In the general case
of arbitrary graph, such an analysis is an open and difficult
problem. Here we discuss some "experimental material" provided by
examples on simple graphs \footnote{The discussion is short since
we plan to devote a separate paper to these examples.}.
\smallskip

In the known examples, in accordance with (\ref{structure of E}),
one encounters
\begin{equation}\label{structure of blocs}
{\mathfrak b}^T_\Phi\,=\,\left\{b \in
C\left(\left[0,\delta\right]; {\mathbb
M}^{M}\right)\,\big|\,\,L_0[b(0)]=0,\,\,L_1\left[b(\delta)\right]=0\right\}\,,
\end{equation}
where $L_0[b(0)]=0$ and $L_1\left[b(\delta)\right]=0$ are the
"linear-type" conditions, which can be (or not be) imposed on
elements $b$. For instance, at $r=0$ there may be
\begin{equation}\label{linear conditions L M}
1-\sum \limits_{i=1}^M b_{i j}(0)=1-\sum \limits_{j=1}^k b_{i
j}(0)=0\,.
\end{equation}
As is seen from (\ref{structure of blocs}), the "massive" part of
the spectrum $\widehat{\mathfrak b}^T_\Phi$ (and
$\widehat{\mathfrak E}^T_\Sigma$: see (\ref{spec E= sum spec b}))
consists of the ideals of the form $${\cal I}_{r_0}\,=\,\{b \in
{\mathfrak b}^T_\Phi\,|\,\,b(r_0)=0\}\,, \qquad r_0 \in (0,
\delta).$$ In the mean time, noncommutativity implies that the
(topologized) spectrum $\widehat{\mathfrak b}^T_\Phi$ is not
necessarily a Hausdorff space: it may contain the clusters. We say
a subset ${\bf c} \subset \widehat{\mathfrak b}^T_\Phi$ to be a
{\it cluster}, if its points are not separable, i.e., for any
point $p\in {\bf c}$ and arbitrary neighborhood $U \ni p$ one has
${\bf c} \subset U$. In examples, clusters do appear as a
consequence of Kirschhoff laws (\ref{Kirchhoff laws}). By them,
the amplitude vectors, which correspond to different families
$\Phi, \Phi^\prime$ are not quite independent. It is the
dependence, which implies the conditions like (\ref{linear
conditions L M}). In the known examples, the clusters in
$\widehat{\mathfrak E}^T_\Sigma$ do appear if $\Omega[\gamma]\cap
\Omega[\gamma^\prime]\ni v$ occurs for the different $\gamma,
\gamma^\prime \in \Sigma$ and an interior vertex $v \in V$.

\subsection{Open questions}
\begin{itemize}
\item By the latter, it would be reasonable to suggest that the
number of interior vertices $n_V$ and the number of clusters
$n_{\bf c}$ are related through an inequality: presumably, $n_V
\geqslant n_{\bf c}$ holds. Since $n_{\bf c}$ is a topological
invariant of spectrum $\widehat{\mathfrak E}^T_\Sigma$, this
relation could be helpful in inverse problems on graphs, in which
the inverse data do determine the eikonal algebra ${\mathfrak
E}^T_\Sigma$ up to an isometric isomorphism (see
\cite{BIP07}--\cite{BW1}). Therefore, the data determine the
spectrum $\widehat{\mathfrak E}^T_\Sigma$ up to a homeomorphism,
and the external observer, which possesses the data, can hope for
getting information about the graph from the spectrum.

\item An intriguing question is whether another geometric
characteristics of the graph (number of edges and cycles,
multiplicity of interior vertices, etc) are related with
topological invariants of the spectrum $\widehat{\mathfrak
E}^T_\Sigma$. A prospective (but rather far) goal is {\it to
recover the graph} from the boundary inverse data via its eikonal
algebra.
\end{itemize}

Hopefully, our approach relates inverse problems on graphs with
C*-algebras. The answers on the above posed questions could
confirm a productivity of these relations.

\end{document}